\documentclass[a4paper,fleqn,usenatbib]{mnras}

\usepackage[T1]{fontenc}
\usepackage{ae,aecompl}

\usepackage{graphicx}	
\usepackage{amsmath}	
\usepackage{amssymb}	
\usepackage{ulem}       
\usepackage{cancel}     
\usepackage{float}
\usepackage{hyperref}
\usepackage{xcolor}
\usepackage{afterpage}
\usepackage{subcaption}
\usepackage{caption}

\def\orcid#1{\kern .08em\href{https://orcid.org/#1}{\includegraphics[keepaspectratio,width=0.7em]{ORCIDiD_icon16x16.png}}}

\title[Non-Jetted TDE State Transitions]{GRRMHD Simulations of State Transitions in Non-Jetted Tidal Disruption Events}
\author[Brandon Curd et al.]{
Brandon Curd$^{1,2,3}$\thanks{E-mail: brandon.curd@utsa.edu}, Safira Heridia$^{1}$, Aviyel Ahiyya$^{1}$, Richard Anantua$^{1,2,3,4
}$
\\ $^{1}$ Department of Physics $\&$ Astronomy, The University of Texas at San Antonio, One UTSA Circle, San Antonio, TX 78249, USA
\\ $^{2}$ Black Hole Initiative at Harvard University, 20 Garden Street, Cambridge, MA 02138, USA
\\ $^{3}$ Center for Astrophysics $\vert$ Harvard \& Smithsonian, 60 Garden Street, Cambridge, MA 02138, USA
\\ $^{4}$ Physics \& Astronomy Department, Rice University, Houston, TX
77005-1827, USA
}
\date{Accepted XXX. Received YYY; in original form ZZZ}

\pubyear{2023}

\begin{document}
\label{firstpage}
\pagerange{\pageref{firstpage}--\pageref{lastpage}}
\maketitle

\begin{abstract}
Circularization of the stream material into a debris cloud during tidal disruption events (TDEs) was recently demonstrated in one of the most accurate long duration TDE simulations to-date. The cooling envelope model (CEM) provides a description of the circularized debris cloud and its emission over time well beyond circularization across different disruption parameters. In the CEM, sub-Eddington accretion rates occur early in TDEs and the debris has a shallow density profile of roughly $\rho \propto r^{-1}$, with Eddington accretion only being achieved after several months. To explore the late stages of the CEM, we perform general relativistic radiation magnetohydrodynamics (GRRMHD) simulations of magnetized tori adapted from the near Eddington phase of the CEM for a $1M_\odot$ star disrupted around a $10^7 M_\odot$ black hole (BH). We find that the disk becomes thermally unstable within 17.1-46.5 days depending on the spin of the BH. Thermal spectra show a soft X-ray excess prior to collapse, with a nearly two order of magnitude decline in X-ray luminosity upon disk collapse. Furthermore, the evolution of the blackbody radius and temperature of our models are correlated with the spin of the black hole. The spectral properties and soft X-ray luminosity in our models are similar to the TDE AT2021ehb, which is a non-jetted TDE with late X-rays and a state transition after $\approx 271$ days.
\end{abstract}




\section{Introduction}

When a star orbits close enough to a black hole (BH) for the gravitational tidal forces to overcome its own self-gravity (a critical radius called the tidal radius, $R_t$), it can be partially or fully disrupted into a stream of bound and unbound gas \citep{Hills1975,Rees1988,Phinney1989,Evans1989}. For the purposes of this work, we assume full disruptions; however, the interested reader may see \cite{2013ApJ...767...25G,2017A&A...600A.124M,2019ApJ...882L..26G,2025ApJ...983..177N} for details on partial disruptions. Within an orbital time of the most bound material, which is nearly the orbital period of the star before disruption and is often referred to as the 'fallback timescale' ($t_{\rm fb}$), the most bound portion of the stream of disrupted gas makes its first return back to pericenter. Apsidal precession causes the gas making its first passage to intersect with gas yet to return. This process leads to shocks, disk formation, and emission. Ultimately, we observe these events as optical, UV, X-ray, and sometimes radio transients with emission lasting up to several years. Crucially, the amount of mass supplied by the stream to pericenter (or the mass fallback rate, $\dot{M}_{\rm fb}$) scales with time as $\dot{M}_{\rm fb}\propto t^{-5/3}$. Since dissipative processes power the observed radiation, either through shocks or viscous accretion, the luminosity in optical/UV/X-ray bands also scales as $L\propto t^{-5/3}$. This has been a hallmark of TDEs and is vital in differentiating them from other transients such as supernovae.

TDEs present themselves observationally with a range of behaviors at peak \citep{Gezari2021,Guolo2024}. Some are only identified in the optical/UV bands initially while others show both optical/UV and X-ray emission near their peak. The optical/UV emission in thermal (or non-jetted) TDEs is characterized by low blackbody temperatures ($T_{\rm BB}\approx \,\rm{ few}\times 10^4$ K) and large photosphere radii ($R_{\rm BB}\approx 10^2-10^4r_g$, depending on the BH mass) which are larger than both the circularization and self-intersection radii. X-ray emission in thermal TDEs on the other hand is hotter ($T_{\rm BB}\approx 10^5-10^6$ K) with small photosphere radii on the order of the horizon scale, sometimes even being smaller than the Schwarszchild radius. A handful (with 3 of 4 cited being jetted events) have shown X-ray state transitions after several months to more than a year \citep{Zauderer2013,Pasham2015,2022ApJ...937....8Y,Eftekhari2024} which may be associated with changes in the accretion state.

\cite{Steinberg2024} demonstrated for the first time via simulations with realistic orbital parameters that Solar mass TDEs around $10^6 M_\odot$ BHs form circularized debris clouds within roughly a fallback time. This is a major result as the properties of the debris cloud are critical to the evolution in a TDE. It is worth noting that \cite{Steinberg2024} exclude magnetic fields; hence, the effects of magnetism during the disk circularization process are not yet known.  We note that several authors have studied the effects of magnetic fields during TDE disk formation, but either in idealized conditions or for shorter durations \citep{Sadowski2016,Curd2021,Meza2025}. We also note that earlier simulations shed light on the dynamics of disk circularization in TDEs, but with less realistic orbital properties due to computational limitations \citep{2016MNRAS.455.2253B,2016MNRAS.461.3760H}. \cite{Metzger2022} presented an analytical model of this debris cloud's evolution, which has been named the cooling envelope model (or CEM, hereafter), which accounts for BH feedback and provides a predictive model across BH mass, stellar mass, and orbital properties of the TDE. \cite{Metzger2022} approximates the cloud as a quasi-spherical distribution, but they only fully describe the fluid properties down to the circularization radius, beyond which some form of accretion flow with a parameterized efficiency is assumed to form, giving an X-ray luminosity proportional to the BH accretion rate $L_X=\eta \dot{M}_\bullet c^2$. Because of the accretion rate in the CEM starting low and increasing over time, the late appearance of X-rays in some TDEs is naturally explained.

Several studies have attempted to explain X-ray and Optical/UV TDEs with the viewing angle dependence in super-Eddington accretion disks \citep{Dai2018,Curd2019,Thomsen2022,Guolo2024}. Since X-rays escape through a narrow funnel with an opening angle of $\sim 10^\circ$, this can explain the prevalence of optical/UV selected TDEs. \cite{Thomsen2022} also demonstrate that the luminosity and photosphere radius show similar evolution to observed TDEs in this model. This model is attractive since the fallback rate is well above Eddington for Solar mass disruptions around BHs of mass $M_\bullet \lesssim 5\times 10^7$. However, the accretion rate is sensitive to the disk assembly process and may not track the fallback rate \citep{Ryu2023}. The mass accretion rate in the CEM, assuming viscous accretion with standard disk viscosity parameters, is not super-Eddington for most of the transient. Instead, the mass accretion starts at only $\sim0.01-0.1$ times the Eddington rate and grows, potentially becoming above Eddington, over time as the debris cloud cools and contracts. For typical BH masses for TDEs, the mass accretion rate in the CEM can take hundreds of days to cross the Eddington limit.

While previous super-Eddington models have demonstrated several important properties, the accretion rate is assumed a priori. That is, the authors start from the assumption of a super-Eddington accretion rate, which is selected based on the fallback rate, not the properties of the debris cloud as in \cite{Metzger2022}. In this work, we provide new simulations which use initial conditions based on the CEM.

We start with the \cite{Metzger2022} CEM and adapt it to a general relativistic magnetohydrodynamics setting with radiation physics incorporated. As the $0.01-0.1$ Eddington range will form a thin accretion disk, it is not affordable with our current code due to extreme angular resolution requirements even using an axisymmetric 2D grid. We instead focus on the stage when accretion for a Schwarzschild BH will be just above Eddington for the CEM. For our assumed BH mass using parameters from \cite{Metzger2022}, the accretion rate from the CEM becomes above Eddington around 250 days after the initial disruption (or 67 days after debris circularization). We simulate the resulting accretion flow of a magnetized torus of gas with initial conditions approximating a debris cloud from the disruption of a star of mass $M_*=1M_\odot$ with orbital impact parameter $\beta=1$ around a Kerr BH of mass $M_\bullet=10^{7} M_\odot$. We simulate both prograde and retrograde systems by setting the unitless BH spin parameter to $a_\bullet = (-0.9, \, 0, 0.9)$.

\section{Numerical Methods} \label{sec:nummethods}

The simulations presented in this work were performed using the general relativistic radiation magnetohydrodynamical (GRRMHD) code \textsc{KORAL}
\citep{2013MNRAS.429.3533S,2014MNRAS.439..503S,2014MNRAS.441.3177M,2015MNRAS.454.2372S,2017MNRAS.466..705S} which solves the conservation equations in a fixed, arbitrary spacetime using finite-difference method. We solve the following conservation equations:
\begin{align}
  (\rho u^\mu)_{;\mu} &= 0, \label{eq:drhou} \\
  (T^\mu_{\ \, \nu})_{;\mu} &= G_\nu, \label{eq:dTmunu} \\
  (R^\mu_{\ \, \nu})_{;\mu} &= -G_\nu, \label{eq:dRmunu} \\
  (n u_R^\mu)_{;\mu} &= \dot{n}, \label{eq:dn} 
\end{align}
where $\rho$ is the gas density in the comoving fluid frame, $u^\mu$ are the components of the gas four-velocity as measured in the ``lab frame'', $T^\mu_\nu$ is the MHD stress-energy tensor in the ``lab frame'':
\begin{equation} \label{eq:eq9}
  T^\mu_\nu = (\rho + u_g+ p_g + b^2)u^\mu u_\nu + (p_g + \dfrac{1}{2}b^2)\delta^\mu_\nu - b^\mu b_\nu,
\end{equation}
$R^\mu_\nu$ is the stress-energy tensor of radiation, $G_\nu$ is the radiative four-force which describes the interaction between gas and radiation \citep{2014MNRAS.439..503S}, and $n$ is the photon number density. Here $u_g$ and $p_g=(\gamma_g - 1)u_g$ are the internal energy and pressure of the gas in the comoving frame (linked by adiabatic index $\gamma_g$) and $b^\mu$ is the magnetic field four-vector which is evolved following the ideal MHD induction equation \citep{2003ApJ...589..444G}. In the \textsc{KORAL} simulations, we assume a single temperature plasma where the ion temperature ($T_i$) and the electron temperature ($T_e$) are identical. This description is adequate in the optically thick regions where collisions are common, but may not be accurate in the extended jet where the gas density is substantially lower \citep{2022ApJ...935L...1L}. We also assume that the electrons follow a thermal distribution.

The radiative stress-energy tensor is obtained from the evolved radiative primitives, i.e. the radiative rest-frame energy density and its four velocity ($u^\mu_R$) following the M1 closure scheme modified by the addition of radiative viscosity \citep{2013MNRAS.429.3533S,2015MNRAS.447...49S}.

The opposite signs of $G_\nu$ in the conservation equations for gas and radiation stress-energy (Eqs. \ref{eq:dTmunu} and \ref{eq:dRmunu}) reflect the fact that the gas-radiation interaction is conservative, i.e. energy and momentum are transferred between gas and radiation, \cite[see][for more details]{2017MNRAS.466..705S}. We include the effects of absorption and emission via the electron scattering opacity ($\kappa_{\rm{es}}$), free-free absorption opacity ($\kappa_{\rm{a}}$), and bound-free absorption opacity through the Sutherland Dopita model \citep{1993ApJS...88..253S} and assume a solar metal abundance for the gas. We also include the effects of thermal synchrotron and Comptonization \citep{2015MNRAS.454.2372S,2017MNRAS.466..705S}. 

Throughout this work, we use gravitational units to describe physical parameters. For distance we use the gravitational radius $r_g\equiv GM_\bullet /c^2$ and for time we use the gravitational time $t_g\equiv GM_\bullet/c^3$, where $M_\bullet$ is the mass of the BH. Often, we set $G=c=1$, so the above relations would be equivalent to $r_g=t_g=M_\bullet$. Occasionally, we restore $G$ and $c$ when we feel it helps to keep track of physical units.

\section{Simulation Details}

Each simulation was evolved in the Kerr-Schild metric with a BH of spin $a_\bullet=(-0.9, \,0, \, 0.9$), with negative spin corresponding to gas in a retrograde orbit around the BH in our setup. Since the simulations presented in this work are conducted in 2D {$r-\vartheta$} coordinates, we implement the mean-field dynamo model described in \citet{2015MNRAS.447...49S} to sustain the magnetic field throughout the simulation.

We set the following parameters for the grid, where we use the same notation as in \citet[][Appendix B]{2017MNRAS.467.3604R}. The hyperexponential break radius is set at $r_{\rm br}=1000r_g$ such that the grid is exponentially spaced at $r<r_{\rm br}$ and becomes hyperexponential at $r>r_{\rm br}$. The collimation radii are $r_{\rm coll,jet}=1000r_{\rm g}$,  $r_{\rm coll,disk}=5r_{\rm +}$, the decollimation radii are $r_{\rm decoll,jet}=r_{\rm decoll,disk}=2r_{\rm +}$. The power-law indices are $\alpha_{\rm 1}=1$, $\alpha_{\rm 2}=0.25$. The fraction of the angular resolution concentrated in the jet and disk are $f_{\rm jet}=0.35$, $f_{\rm disk}=0.4$. We 'cylindrify' cells close to the axis at small radii by expanding their size in $\vartheta$ \citep{2011MNRAS.418L..79T,2017MNRAS.467.3604R}; the cylindrification radius $r_{\rm cyl}=20r_{\rm g}$ and $n_{\rm cyl}=1$. This increases the minimum time step of the simulation and increases the computational efficiency by nearly 2-3 times. The polar angle code coordinate $x_2$ extends from
$x_{\rm 2,min}=10^{-5}$ to $x_{\rm 2,max}=1-10^{-5}$, where $x_2=0$ and 1 correspond to the two polar axes.

We set the resolution $N_r \times N_\vartheta=640\times480$ such that the cells are roughly 1:1 in most of the simulation domain except beyond $r>r_{\rm br}$. We use modified Kerr-Schild coordinates with the inner edge of the domain inside the BH horizon. At the inner radial boundary ($R_{\rm{min}}=0.85r_H$), we use an outflow condition while at the outer boundary ($R_{\rm{max}}=10^5r_g$), we use a similar boundary condition and in addition prevent the inflow of gas and radiation. Our choice of $R_{\rm{min}}$ is such that 9 cells in the computational domain lie inside of the horizon. At the polar boundaries, we use a reflective boundary. The inner two cells at the jet axis are not evolved and instead we copy the primitives from the 3rd cell from the polar axis to bolster numerical stability. To maintain numerical stability, we introduce mass in highly magnetized regions of the simulation domain using a floor condition on the magnetization $\sigma \equiv b^2/\rho \leq 100$ throughout each simulation.

\begin{figure}
    \centering{}
    \includegraphics[width=\columnwidth]{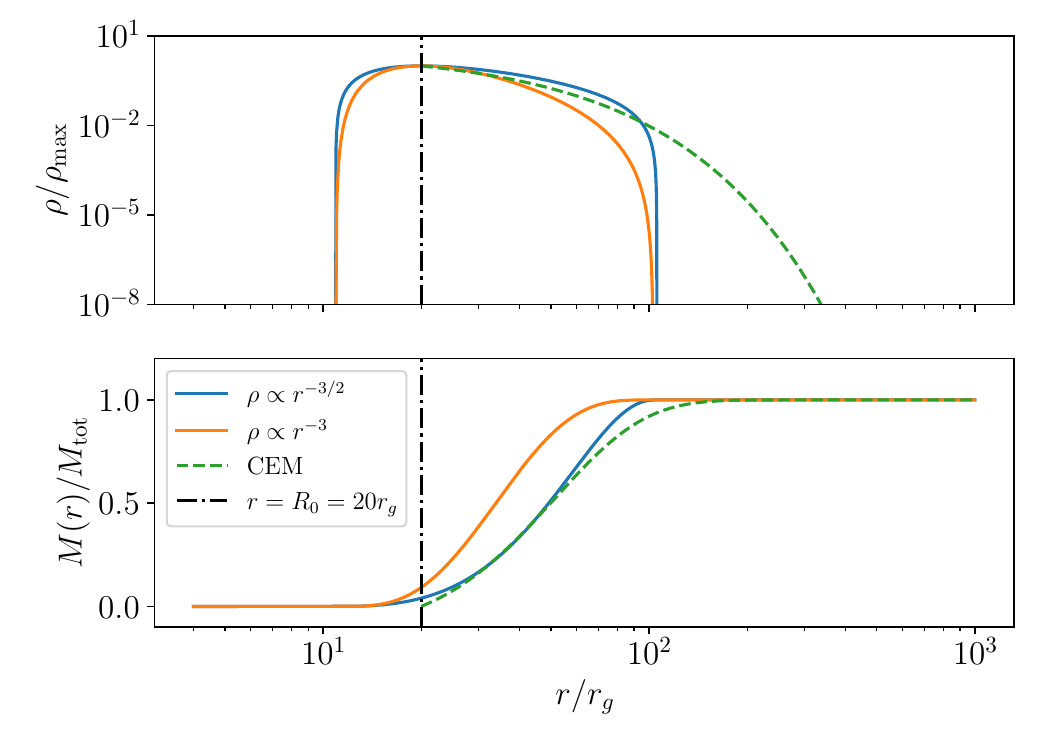}
    \caption{\textcolor{black}{Here we compare the mid-plane density (top) and integrated mass (bottom) as a function of radius for our initial conditions (blue), the Kato torus model used in \protect\cite{Curd2019} but with a similar torus size (orange), and the CEM model used to fix the Kato torus parameters used in this work (green). We also indicate the location of the density maximum ($R_0$, dash-dotted line). The CEM curve is cutoff at $r<R_0$ since inside of this region, some accretion flow is assumed to form that has an unspecified density profile. The Kato torus model with $\rho \propto r^{-3/2}$ has a similar integrated mass and density profile up to $\sim100 r_g$, beyond which the torus model sharply cuts off. We note the CEM has a slower decline after $100r_g$, but the total mass beyond this radius is negligible and the model used minimizes the difference in both density and integrated mass.}}
    \label{fig:initial_density}
\end{figure}

As a starting point for the debris cloud, we use the hydrostatic torus presented by \cite{Kato2004}. Details on how we implement the \cite{Kato2004} model in KORAL are given in Appendix A in \cite{Curd2019}. We assume a constant angular momentum torus. The density maximum ($R_0$) is set to the circularization radius $R_{\rm{circ}}=2R_t\approx 20 r_g$ for a $\beta=1$ TDE of a Solar mass star. We set the angular momentum to $l_0=\sqrt{GM_\bullet R_{\rm{circ}}^3}/(R_{\rm{circ}}-r_S)$, where $r_S=2GM_\bullet/c^2$ is the Schwarzschild radius. A major difference between this work and \cite{Curd2019} is that we fix the initial density profile using the assumed profile from \cite{Metzger2022}, which is much shallower than our previous model \textcolor{black}{(see \autoref{fig:initial_density})}. The virial radius defining the profile is $R_v=17.2r_g$, which is inside the circularization radius so the initial cloud is primarily determined by an exponential decline. \textcolor{black}{Since the disk will be radiation pressure dominated at the initial accretion rate for the CEM solution, we set the adiabatic index to $\Gamma=4/3$.} Once we obtain the initial hydrodynamic solution giving $p_{\rm gas,0}$ \textcolor{black}{(determined by fitting to the CEM solution)}, we split the gas pressure into gas and radiation pressure by solving $p_{\rm gas,0}=p_{\rm gas}+p_{\rm rad}$. We list the relevant quantities of each model in \autoref{tab1}.

To initialize the magnetic field, we define the vector potential
\begin{equation}
    A_\varphi \propto \max (\rho-0.1\rho_{\rm max},0),
\end{equation}
where $\rho_{\rm max}$ is the gas density at $R_{\rm circ}$. The magnetic field strength is then scaled such that the maximum pressure ratio is $\beta_{\rm mag}\equiv (p_{\rm gas}+p_{\rm rad})/p_{\rm mag}=100$. This choice is made in order to minimally resolve the magnetorotational instability (MRI) in the disk upon initialization, but such values of $\beta_{\rm mag}$ are likely to be realized in a TDE disk within a few orbital times due to a dynamo effect \citep{Sadowski2016,Curd2021,2026ApJ..1001...71A}.

In each simulation, we initially start with a torus mass 100 times the target initial mass to improve stability until gas from the torus begins to accrete. After $\sim2t_{\rm visc}\approx 20 (R_{\rm circ}/r_g)^{3/2}t_g$, the accretion rate settles to its maximum. We then rescale the gas density at $t_{\rm rescale}=15,000 t_g$ such that the total mass in the simulation domain is $\sim0.14M_\odot$ to match the CEM model for a TDE around a Schwarzschild BH \citep{Metzger2022} and continue the simulation until the disk becomes thermally unstable. The simulation time since rescaling ($\Delta t = t_{\rm sim} - t_{\rm rescale}$) is used as the physically relevant time.

\begin{table}
    \centering
    \begin{tabular}{ l c c c c}
        \hline
        \hline 
        Model & $M_\bullet$ & $a_\bullet$ & $\dot{M}_{\bullet, {\rm max}}$& $\Delta t_{\rm final}$ \\
                & ($M_\odot$) & & ($\rm{g \ s^{-1}}$) & (days) \\
        \hline
        \texttt{m7am.9-M22} & $10^7$ & -0.9 & $7.6\times10^{26}$ & 17.1 ($30,000t_g$) \\
        \texttt{m7a0-M22} & $10^7$ & 0 & $1.1\times10^{26}$ & 31.3 ($55,000t_g$) \\
        \texttt{m7a0.9-M22} & $10^7$ & 0.9 & $3\times10^{25}$ & 46.4 ($81,500t_g$) \\
    \hline
    \end{tabular}
    \caption{Model parameters.}
    \label{tab1}
\end{table}

\section{Definitions} \label{sec:definitions}

In this section, we discuss the units adopted throughout the text and provide brief descriptions of quantities used to study the \textsc{KORAL} simulation data. 

We adopt the following definition for the Eddington mass accretion rate:
\begin{equation} \label{eq:mdotEdd}
  \dot{M}_{\rm{Edd}} = \dfrac{L_{\rm{Edd}}}{\eta_{\rm NT} c^2},
\end{equation}
where $L_{\rm{Edd}} = 1.25\times 10^{38}\, (M_\bullet/M_\odot)\, {\rm erg\,s^{-1}}$ is the Eddington luminosity, $\eta_{\rm{NT}}$ is the radiative efficiency of a thin disk around a BH with spin parameter $a_\bullet$ (which is often referred to as the Novikov-Thorne efficiency):
\begin{equation} \label{eq:etaNT}
  \eta_{\rm{NT}} = 1 - \sqrt{1 - \dfrac{2}{3 r_{\rm{ISCO}}}},
\end{equation}
and $r_{\rm{ISCO}}=3+Z_2 - \sqrt{(3-Z_1)(3+Z_1+2Z_2)}$ is the radius of the Innermost Stable Circular Orbit (ISCO, \citealt{1973blho.conf..343N}) in the Kerr metric, where $Z_1 = 1 + (1-a_\bullet^2)^{1/3}\left((1+a_\bullet)^{1/3}+(1-a_\bullet)^{1/3}\right)$ and $Z_2 = \sqrt{3a_\bullet^2 + Z_1^2}$. For $a_\bullet =$ 0, the efficiency is $\eta_{\rm{NT}}=$ 5.72\%.

We compute the net mass inflow rate as
\begin{equation} \label{eq:mdotin}
  \dot{M}(r) = -2\pi\int_0^\pi\sqrt{-g}\rho \,u^r d\vartheta .
\end{equation}
Similarly, we take the net mass outflow rate as
\begin{equation} \label{eq:mdotin}
  \dot{M}_{\rm out}(r) = 2\pi \int_0^\pi\sqrt{-g}\rho \, \max{(u^r,0)} d\vartheta .
\end{equation}
\textcolor{black}{When calculating the mass outflow rate, we calculate it at $50r_g$ to avoid adding any contamination from regions where the initial conditions have not yet reached outflow equilibrium and we also only consider unbound gas.}

The magnetic flux is computed as
\begin{equation} \label{eq:magflux}
  \Phi(r) = -\pi \int_0^{\pi}\sqrt{-g}|B^r(r)|d\vartheta ,
\end{equation}
where $B^r$ is the radial component of the magnetic field. 

The total energy flux (the net energy flux including radiation) is computed as:
\begin{equation} \label{eq:ltot}
  L_{\rm{net}}(r) = -2\pi\int_{0}^{\pi}\sqrt{-g} (T^r_{\ \, t} + R^r_{\ \, t} + \rho u^r) d\vartheta,
\end{equation}
where we integrate the radial flux of energy carried by gas plus magnetic field ($T^r_{\ \,t}$) and radiation ($R^r_{\ \, t}$), and subtract out the rest-mass energy ($\rho u^r$) since it does not lead to observational consequences for an observer at infinity. The radiative luminosity is given by:
\begin{equation} \label{eq:eq16}
  L_{\rm{rad}}(r) = -2\pi \int_{0}^{\pi}\sqrt{-g} R^r_{\ \, t} d\vartheta,
\end{equation}
which gives the flux of radiation energy through a surface at a given radius. In this work, we measure both $\dot{M}_{\rm out}$ and $L_{\rm{rad}}$ through a sphere at $r_{\rm out}=500\,r_g$. This lies beyond the outer radius of the photosphere. We assume rays crossing the surface reach a distant observer. 

To characterize the disk and corona, we compute several weighted quantities at a fixed radius as 
\begin{equation} \label{eq:weightedint}
  \langle X \rangle_w (r) = \dfrac{\int_0^\pi w(r,\vartheta,\varphi) X(r,\vartheta,\varphi) \sqrt{-g}d\vartheta }{\int_0^\pi w(r,\vartheta,\varphi) \sqrt{-g}d\vartheta },
\end{equation}
where $w$ is the weight, and $X$ is some fluid quantity. 

We estimate the electron scattering photosphere location for an observer at infinity along the direction ($\vartheta$) by integrating the optical depth radially inward from the outer boundary of the grid. We simply integrate at constant ($\vartheta$) in the ``lab frame'' and ignore any effects of curvature:
\begin{equation} \label{eq:taures}
  \tau_{\rm{r}}(r) = \int_r^{R_{\rm{max}}} \rho \kappa_{\rm{es}} \sqrt{g_{rr}}dr',
\end{equation}
where $\kappa_{\rm{es}} = 0.2(1+X)\,{\rm cm^2 \, g^{-1}}$ is the electron scattering opacity, $X$ is the hydrogen mass-fraction which is assumed to be the Solar abundance $X_\odot=0.7381$, and $R_{\rm{max}}$ is the radius corresponding to the outer boundary of the grid. It is also useful to calculate the angle integrated photosphere (with $r$ held constant)
\begin{equation} \label{eq:tauthetaes}
  \tau_{\rm{\vartheta}}(\vartheta) = \int_0^{\vartheta} \rho \kappa_{\rm{es}} \sqrt{g_{\vartheta\vartheta}}d\vartheta',
\end{equation}
For the gas and radiation temperatures in the simulations presented here, the Klein-Nishina correction is negligible, and the electron scattering opacity is essentially the Thomson opacity. We choose the location of the photosphere for radial (angular) integrated case as the $\tau_{\rm{r}}=1$ ($\tau_\vartheta=1$) surface.

\begin{figure*}
    \centering{}
    \includegraphics[width=0.24\textwidth]{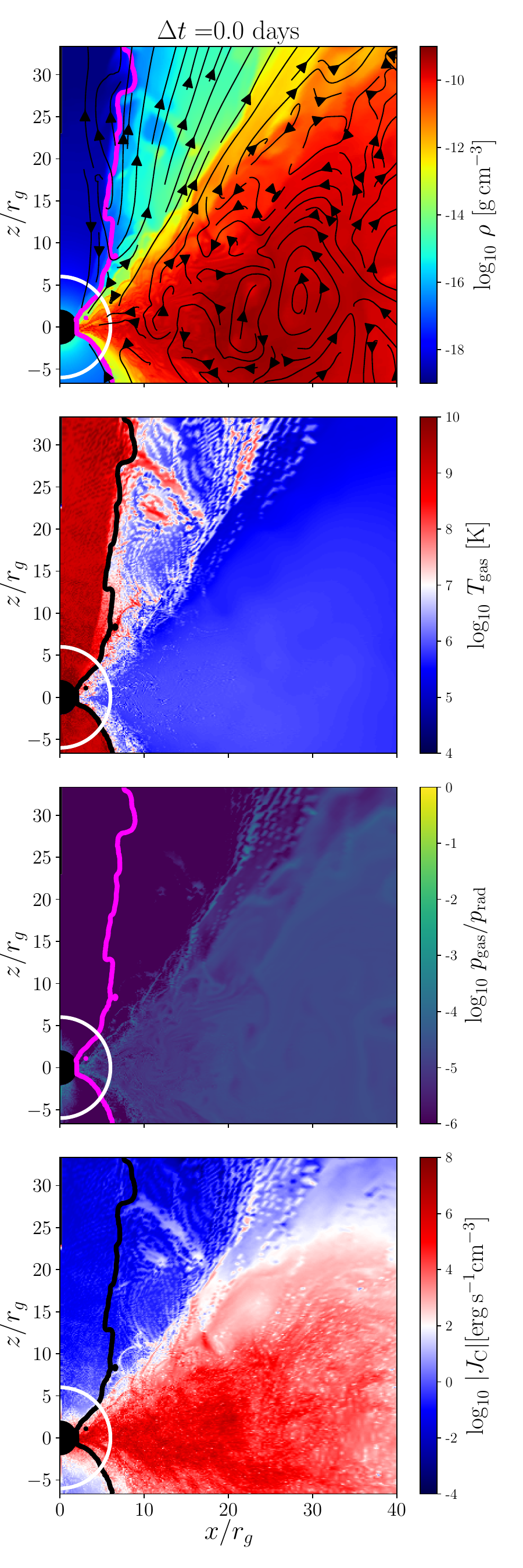}
	\includegraphics[width=0.24\textwidth]{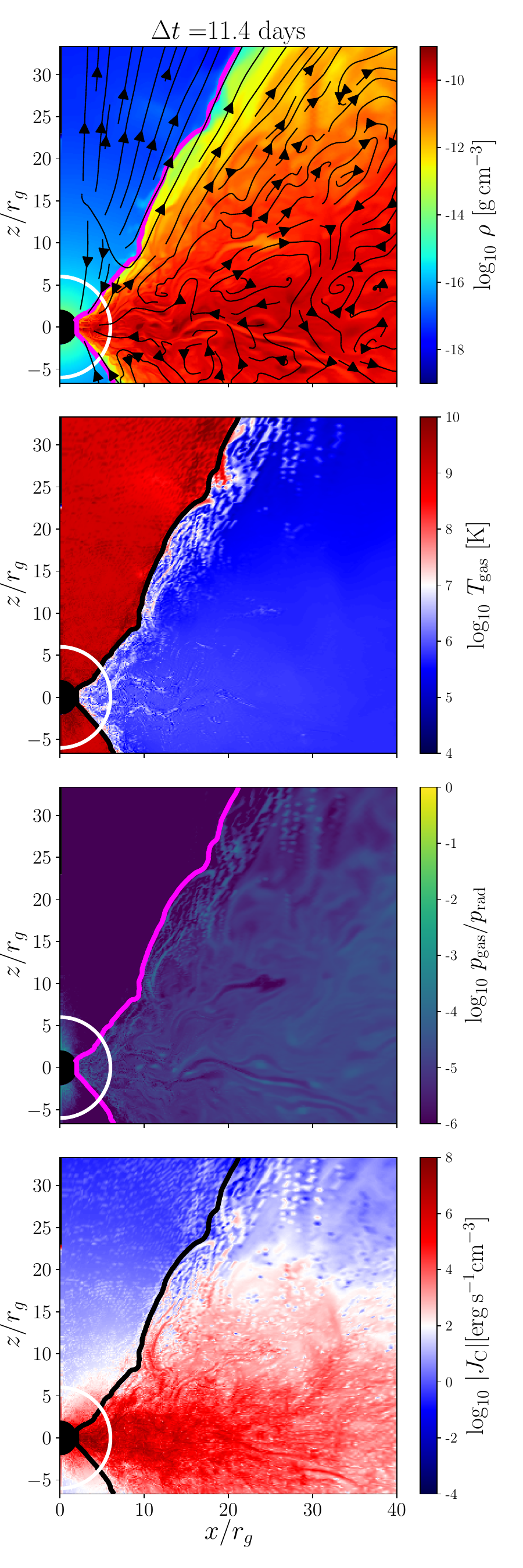}
	\includegraphics[width=0.24\textwidth]{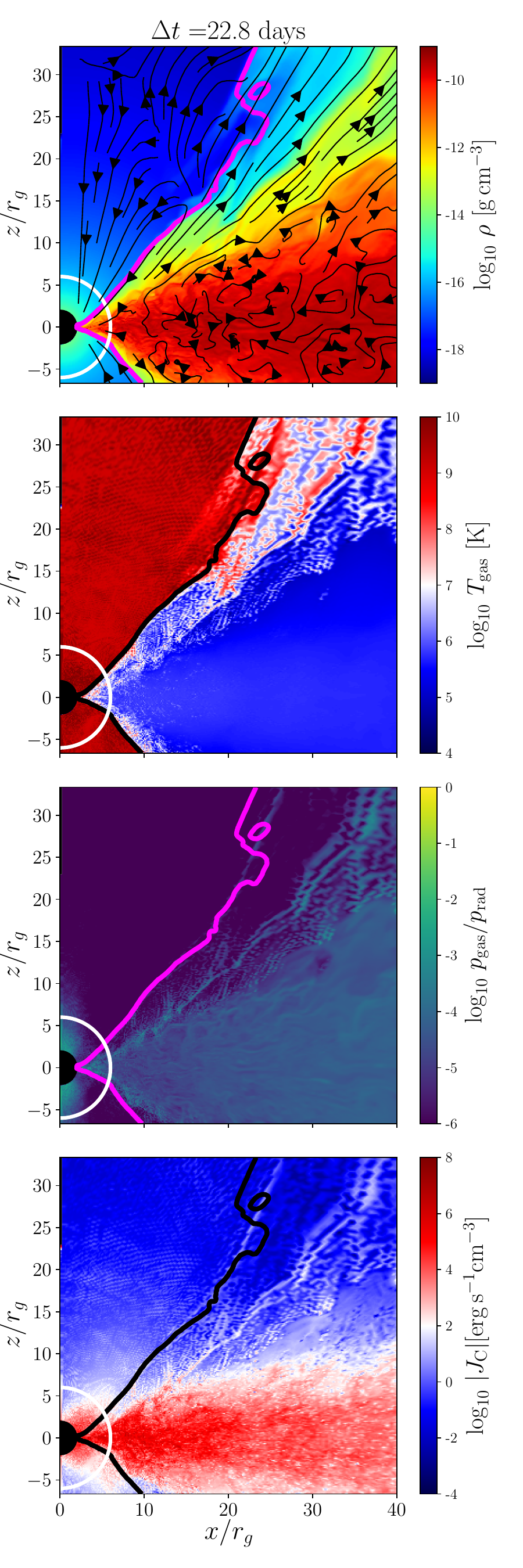}
	\includegraphics[width=0.24\textwidth]{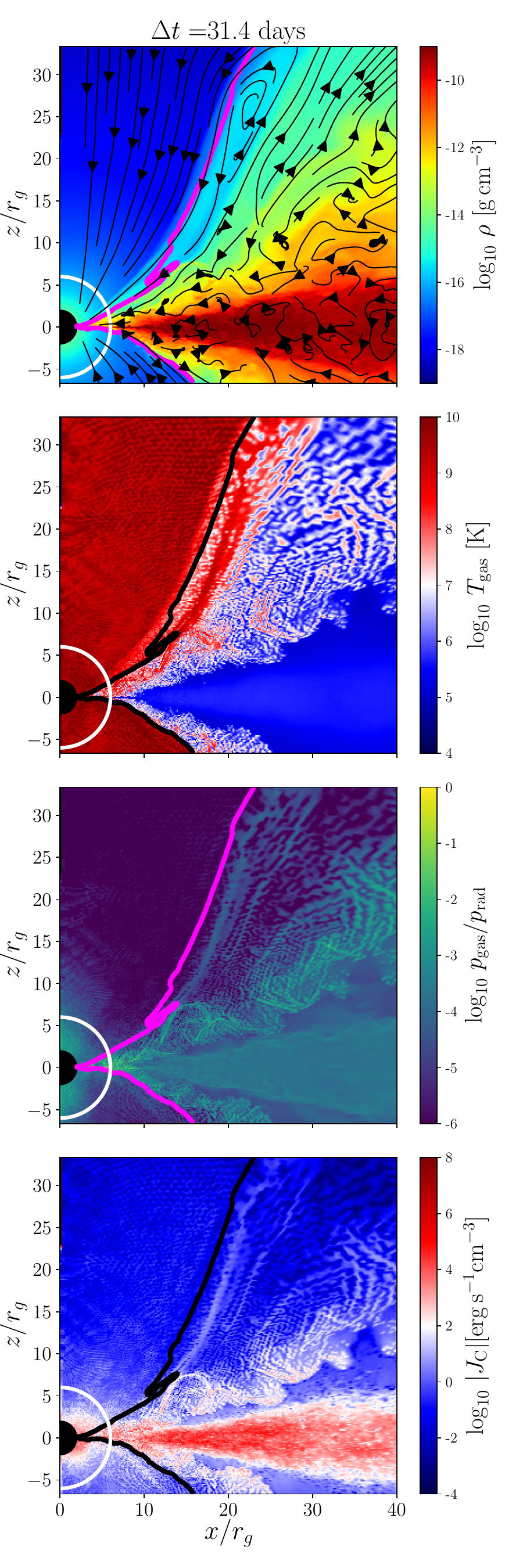}
    \caption{A zoomed in view of the accretion flow and funnel of model \texttt{m7a0-M22} at $\Delta t=$ 0, 11.4, 22.8, and 31.4 days (increasing from left to right). The colors show the gas density (1st row), gas temperature (2nd row), gas to radiation pressure ratio (3rd row), and the Compton cooling rate (4th row). Note that the Compton cooling rate, $J_C$, is estimated following expressions from \protect\cite{2015MNRAS.454.2372S}. Streamlines indicate the fluid velocity. Solid lines (black and magenta) indicate the $\sigma=1$ boundary separating the disk/wind from the low density, magnetized corona. The ISCO radius is indicated as the white circle. The disk height visibly decreases over time. We also note that the gas temperature outside of the disk increases due to Compton cooling becoming less efficient as the accretion rate declines. After the disk collapses (rightmost panel), the pressure ratio approaches unity and the disk appears to truncate near the ISCO radius as can be seen by the drastic decrease in density inside the ISCO radius.}
    \label{fig:heatmaps_zoom}
\end{figure*}

We track the time evolution of the mass accretion rate, magnetic flux, and jet power through unitless quantities evaluated at the BH horizon. We track the accretion of mass onto the BH
\begin{equation} \label{eq:mdotin}
  \dot{M}_\bullet = \dot{M}(r_H).
\end{equation}
We quantify the magnetic field strength at the BH horizon through the normalized magnetic flux parameter \citep{2011MNRAS.418L..79T}
\begin{equation} \label{eq:eq20}
  \phi = \dfrac{\Phi(r_H)}{\sqrt{\dot{M}(r_H)}}.
\end{equation} 
For geometrically thick disks the magnetically arrested disk (MAD, \citealt{Narayan2003}) state is achieved once $\phi\sim 40-50$ \citep[see e.g.][]{2011MNRAS.418L..79T,2012JPhCS.372a2040T}. Meanwhile, weakly magnetized disks which transport angular momentum via MRI turbulence and have $\phi \sim 5$ are typically referred to as standard and normal evolution (SANE) disks in the simulation literature. This class of disks is thought to apply to thermal TDEs \citep{Curd2019} and is the class of disks studied in this work. Lastly, we quantify the total efficiency (sometimes called the `wind' efficiency)
\begin{equation} \label{eq:etaout}
  \eta_{\rm tot} = \dfrac{L_{\rm{net}}(5r_g)}{\dot{M}_\bullet c^2}.
\end{equation} 

We estimate the disk temperature using the density-weighted gas temperature,
\begin{equation} \label{eq:Tdisk}
    \langle T_{\rm disk}\rangle_{15r_g}=\langle T \rangle _\rho (15r_g).
\end{equation}

Finally, we obtain an estimate for the radiation temperature (giving a representation of the typical photon energy) from the radiation energy density
\begin{equation} \label{eq:Trad}
   T_{\rm rad}=\left(\dfrac{\widehat{E}}{a}\right)^{1/4},
\end{equation}
\textcolor{black}{where $a$ is the radiation constant.}

\begin{figure*}
    \centering{}
    \includegraphics[width=\textwidth]{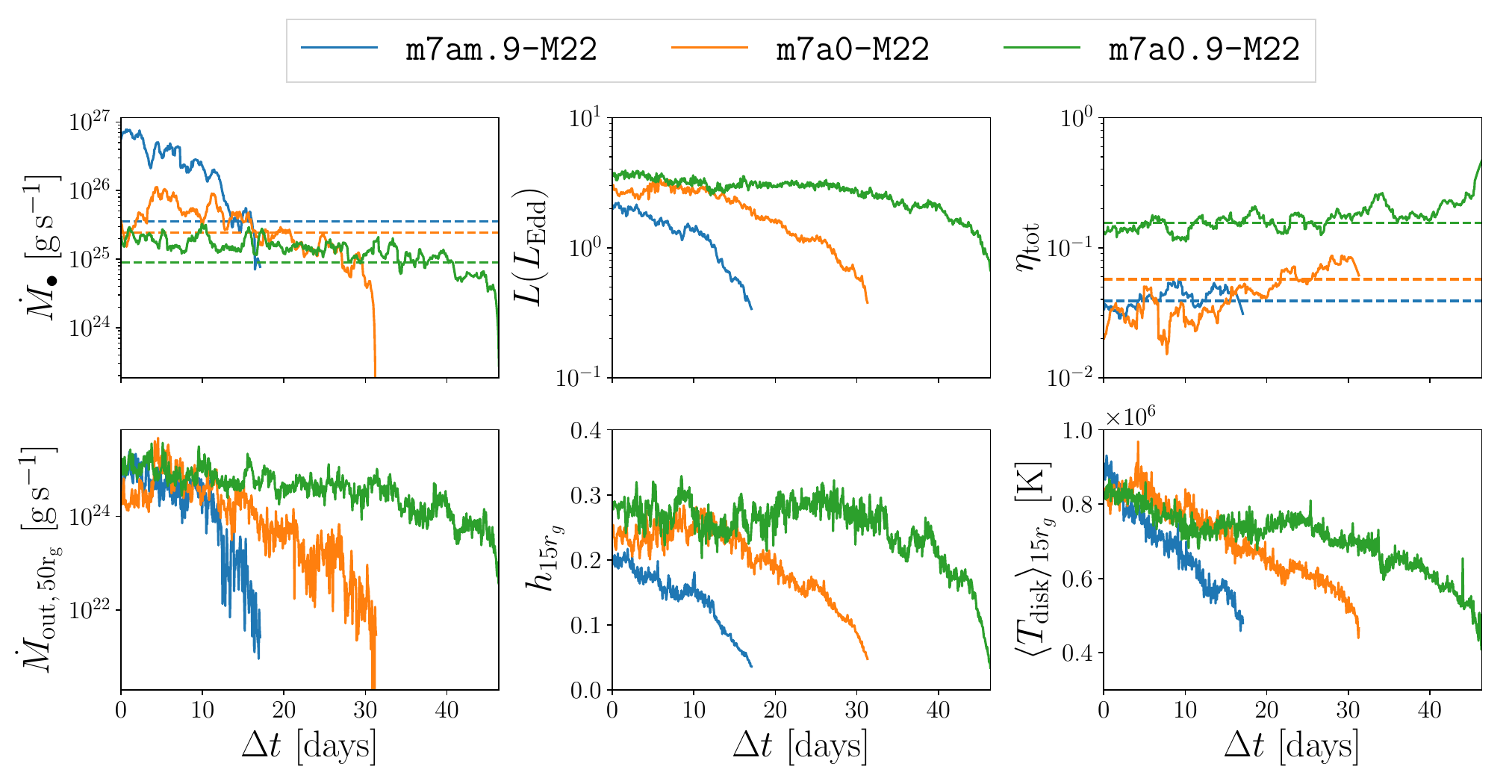}
    \caption{\textcolor{black}{Here we show the mass accretion rate (top, 1st column), luminosity (top, 2nd column), total efficiency (top, 3rd column), mass outflow rate of unbound gas at $50 r_g$ (bottom, 1st column), density scale height (bottom, 2nd column), and density weighted disk temperature (bottom, 3rd column). In the top left panel, we show the Eddington mass accretion rate (horizontal dashed lines) for comparison. In the top right panel, we also show the Novikov-Thorne thin disk efficiency for each spin for comparison (horizontal dashed lines).} Models show a correlation between density scale height and disk temperature as expected. We also note that the retrograde ($a_\bullet=-0.9$) model collapses first, possibly due to weaker feedback (likely some fraction of $\eta_{\rm tot}$). Meanwhile the prograde ($a_\bullet=0.9$) model collapses last. In physical units, the retrograde (prograde) model would go through a state transition $\sim15$ days earlier (later) than the non-spinning model. We apply a 1st order smoothing to the accretion rate and efficiency curves for legibility.}
    \label{fig:scalars}
\end{figure*}

\begin{figure}
    \centering{}
    \includegraphics[width=\columnwidth]{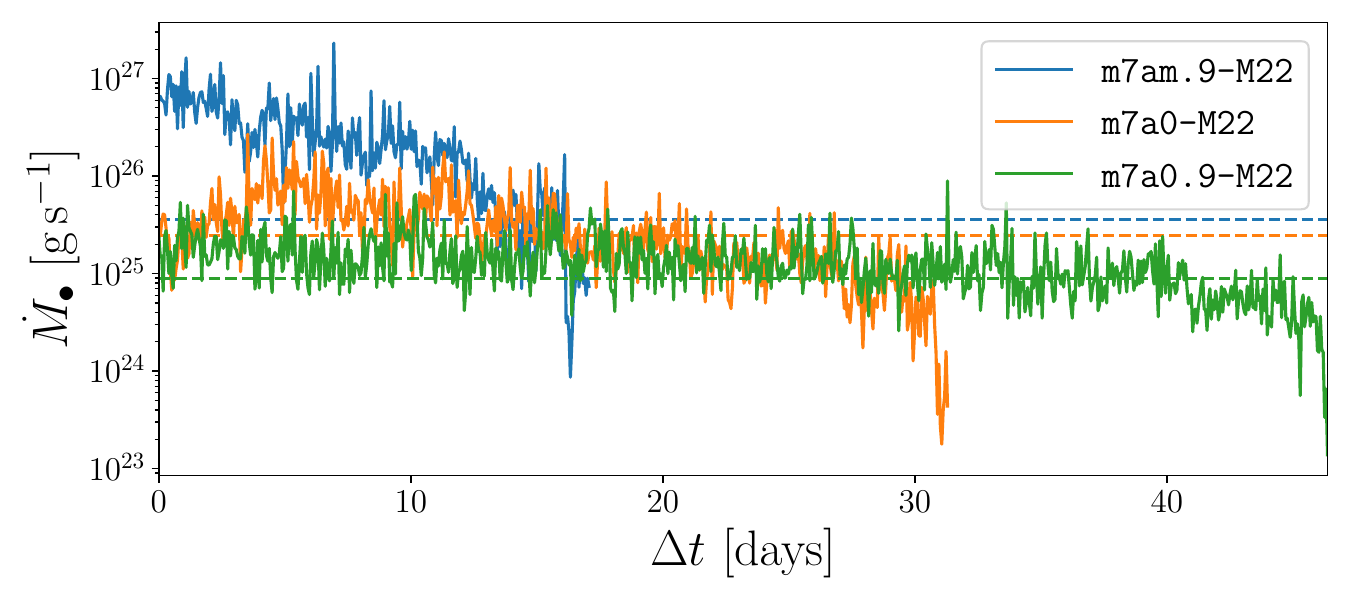}
    \caption{Here we show the mass accretion rate without smoothing.}
    \label{fig:scalars_mdot}
\end{figure}

\begin{figure}
    \centering{}
    \includegraphics[width=\columnwidth]{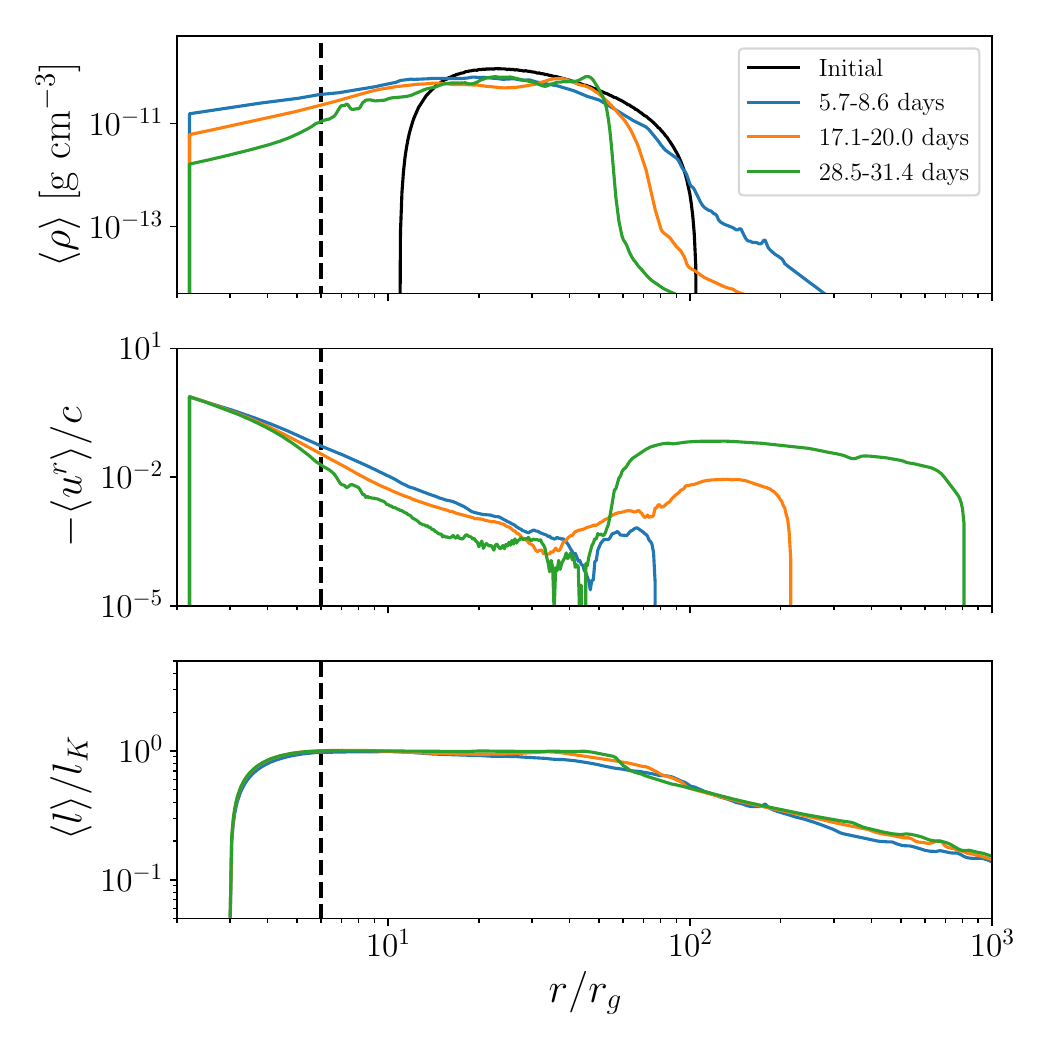}
    \caption{Here we show the radial profiles of gas density (top), radial inflow velocity (middle), and angular momentum normalized by the Keplerian angular momentum (bottom) for model \texttt{m7a0-M22}. The ISCO radius is indicated with the vertical dashed line. Note that quantities are time averaged over chunks of $5,000t_g$ (or 2.8 days).}
    \label{fig:radial_profiles_m7a0}
\end{figure}

\begin{figure}
    \centering{}
    \includegraphics[width=\columnwidth]{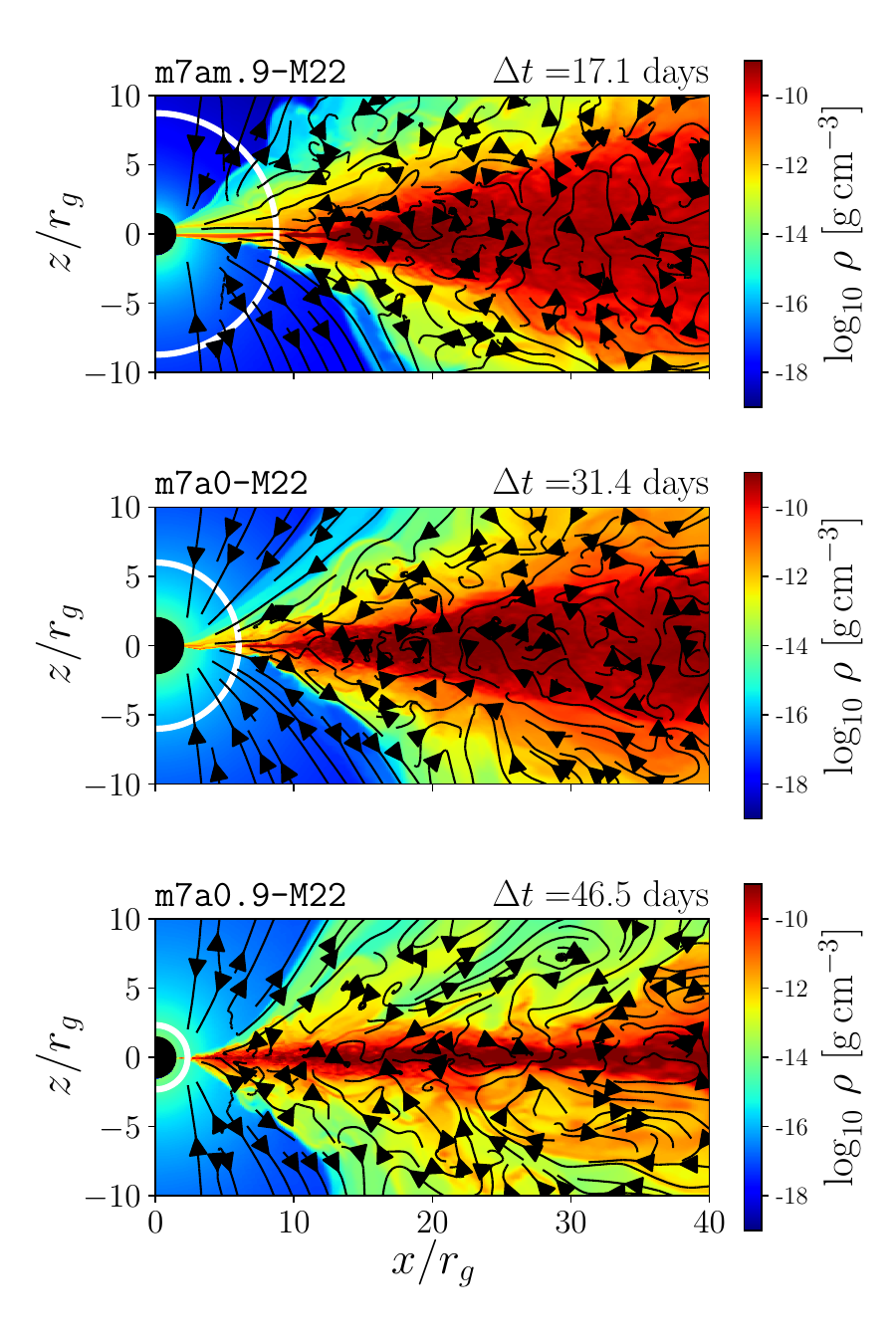}
    \caption{Here we show the accretion disk after thermal collapse for each model. The colors show the gas density, streamlines indicate the fluid velocity, and the ISCO radius is indicated as the white circle. The disk is observed to truncate near the ISCO in each model.}
    \label{fig:heatmaps_ISCOs}
\end{figure}

\begin{figure}
    \centering{}
    \includegraphics[width=0.49\columnwidth]{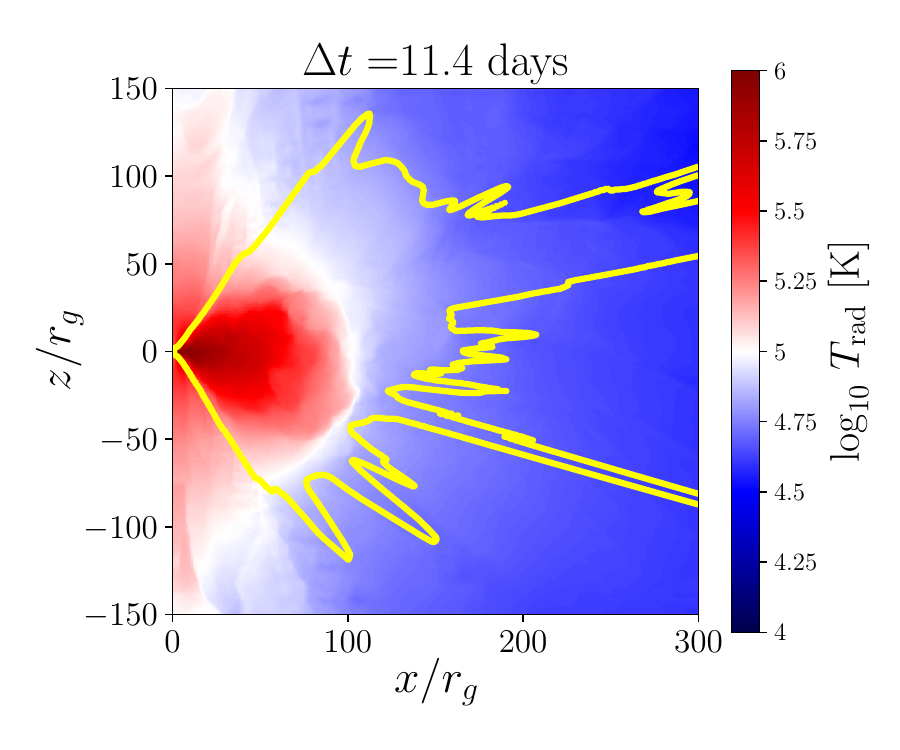}
    \includegraphics[width=0.49\columnwidth]{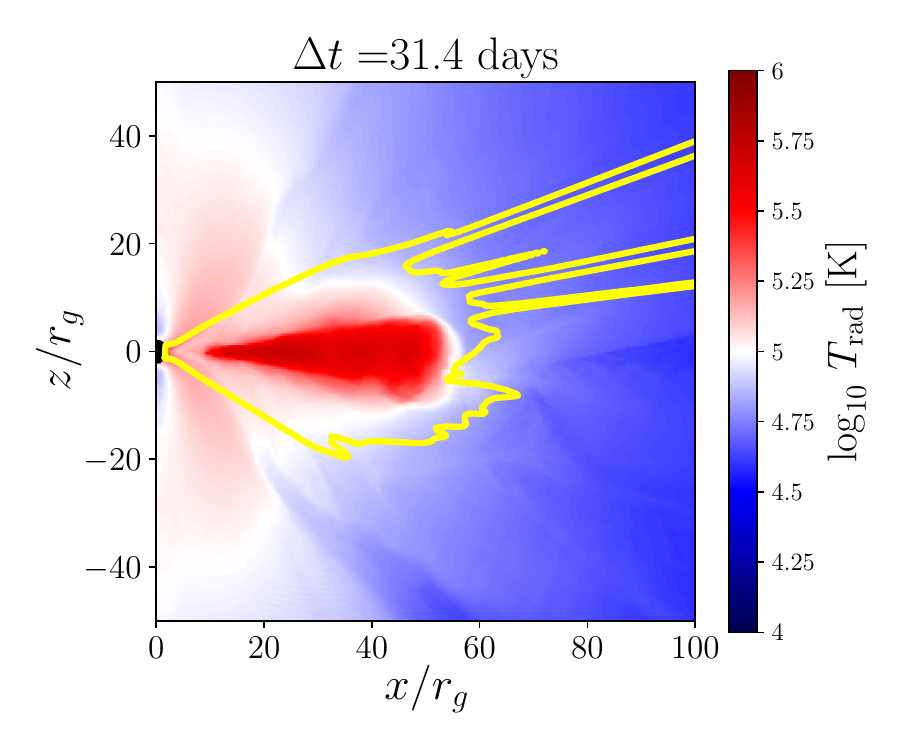}\\
    \includegraphics[width=0.49\columnwidth]{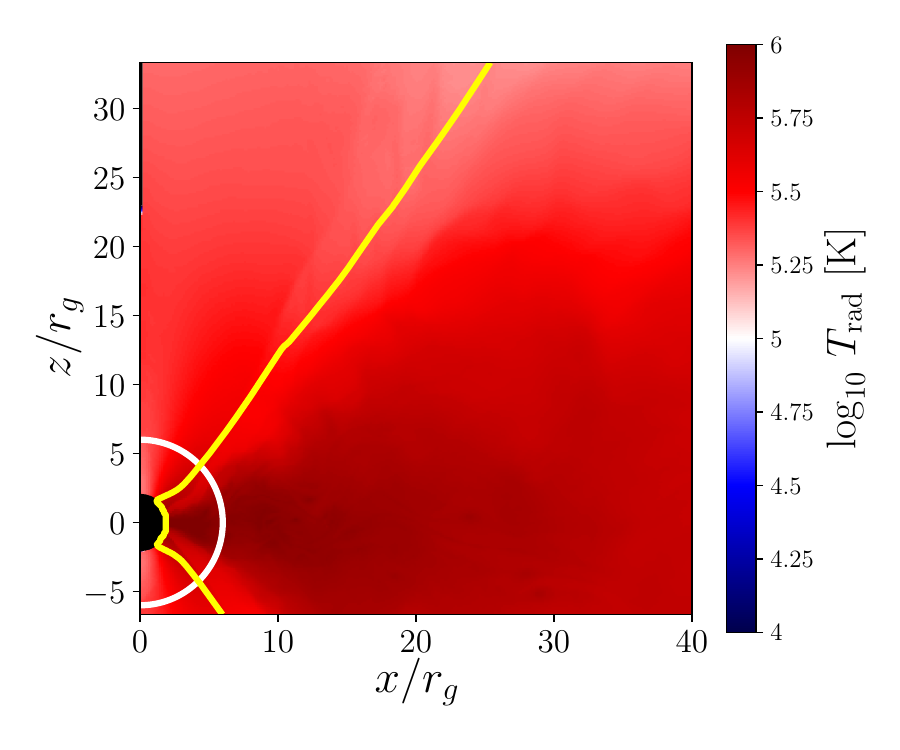}
    \includegraphics[width=0.49\columnwidth]{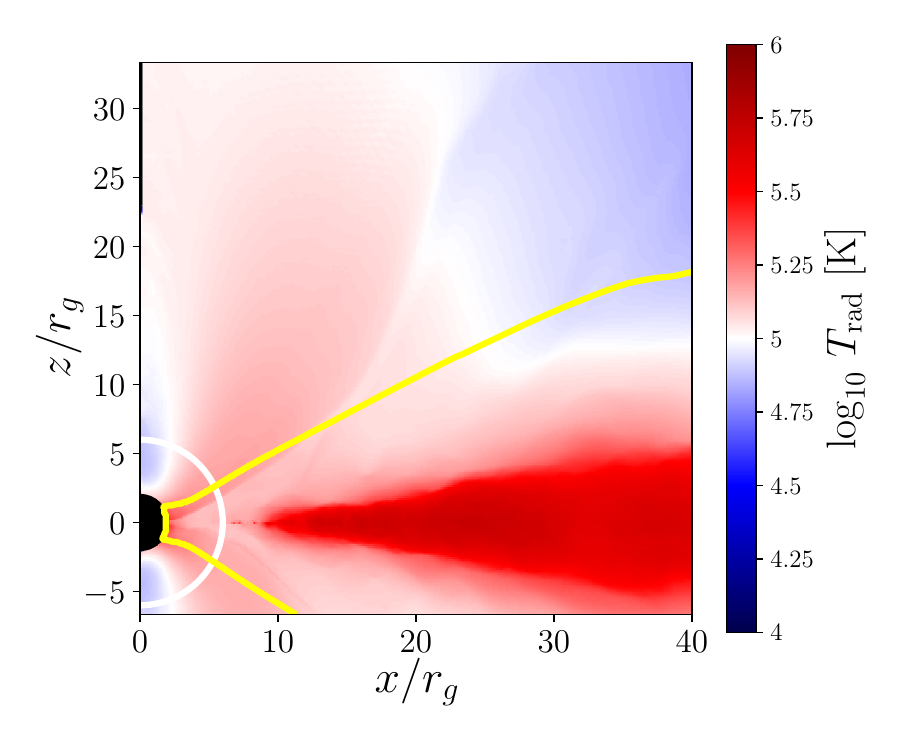}
    \caption{Here we show the radiation temperature (colors), radially integrated photosphere location (yellow line), and ISCO (white circle) for model \texttt{m7a0-M22}. The top panels shows a zoomed out view at 11.4 days (left) and 31.4 days (right). Note that different ranges for the coordinates are used to emphasize the temperature at the outer photosphere surface. The bottom panels show the same times but zoomed in to highlight the radiation temperature in the funnel near the BH horizon where X-rays originate. Note the drastic change in $T_{\rm rad}$ in the funnel after the disk collapses. This visually confirms our thermal spectra results which show an X-ray state transition.}
    \label{fig:heatmaps_Trad}
\end{figure}

\section{Results}

The run time and maximum achieved accretion rate for each simulation are shown in \autoref{tab1}. The spin $-0.9$, 0, and $0.9$ models were each run for $30,000 t_g$ (17.1 days), $55,000 t_g$ (31.4 days), and $81,500 t_g$ (46.5 days), respectively. We verify that each model is SANE throughout the entire evolution as the average normalized magnetic flux at the horizon over the entire duration of the simulation is $\langle \phi \rangle \approx$ 2.2, 3.3, and 5.5 for models \texttt{m7am.9-M22}, \texttt{m7a0-M22}, and \texttt{m7a0.9-M22}, respectively. 

\subsection{Simulation Properties}

The general evolution of the disk in each model is depicted using model \texttt{m7a0-M22} ($a_\bullet=0$) in \autoref{fig:heatmaps_zoom}. At $\Delta t=0$, the disk is geometrically thick and the mass accretion rate is near/above Eddington. A cold ($T\approx 10^{6}$ K) disk is sandwiched between a hot ($T>10^8$ K), magnetized corona (or funnel). Radiation pressure dominates in the optically thick disk. As is typical of thick accretion disks, no edge is observed at the ISCO radius. As the disk evolves, it becomes more gas pressure dominated until it ultimately collapses into a thin disk. At the final time ($\Delta t=31.4$ days), the disk is vertically thinner and the gas temperature has cooled to roughly half of its initial value. The inner disk, especially near the ISCO, is roughly equally supported by gas and radiation pressure. Furthermore, the disk is truncated at the ISCO. This evolution is seen in all models, but with different evolutionary timescales towards collapse.

The time evolution of various diagnostics for each model are shown in \autoref{fig:scalars}. In \autoref{fig:scalars_mdot}, we show an unsmoothed mass accretion rate for completeness. Radial profiles for the gas density, radial inflow velocity, and angular momentum are shown in \autoref{fig:radial_profiles_m7a0} for model \texttt{m7a0-M22}. The spin 0 and 0.9 models show near-Eddington mass accretion early on. The spin -0.9 model clearly accretes much faster initially, reaching above 10 times the Eddington rate. This result qualitatively agrees with the study of super-Eddington disks by \cite{Utsumi2022}, who similarly initialized each of their models with the same torus mass. An in depth analysis to determine the root cause of the enhanced accretion rate is beyond the scope of this work. However, it is likely due to the weaker, less energetic mass outflow for $a_\bullet =-0.9$. Accretion declines over time instead of increasing as the CEM predicts because radial contraction of the disk does not drive the density near the BH up (see \autoref{fig:radial_profiles_m7a0}). Instead, the density near the BH slowly declines as the BH consumes the disk.

The luminosity shows a correlation with spin, as the -0.9 spin model is the lowest and the 0.9 spin model is the highest. Also, the decline in $L$ over time is faster in the retrograde case since $L$ is correlated with $\dot{M}_\bullet$. The efficiency in the spin 0 model is much lower than $\eta_{\rm NT}$ initially, but later slightly exceeds it as the disk becomes thinner. The efficiency of the -0.9 and 0.9 models oscillate about the NT efficiency ($\eta_{\rm NT}$) throughout the entire evolution. Additional power in the form of Poynting flux is present in high spin models \citep[e.g.][]{Utsumi2022} which, even during the radiatively inefficient stages, allows the total efficiency to achieve the NT power, which is derived without accounting for magnetism.

The disk scale height and disk temperature steadily decrease, reflecting the visualization in \autoref{fig:heatmaps_zoom}. Conversely, the corona temperature increases over time as the disk cools and collapses. This is due to the inverse Compton cooling rate declining and then rapidly dropping off as the disk cools and collapses. We are forced to stop each simulation when the disk becomes under-resolved (fewer than $\sim10$ cells per density scale height). After collapse, each model is observed to truncate near the ISCO radius \autoref{fig:heatmaps_ISCOs}. 

The radial inflow velocity indicates that the inflow speed decreases as the disk collapses, which is in agreement with predictions for the behavior of disks as they transition from slim to thin disk solutions. The gas is generally roughly Keplerian throughout the bulk of the disk.

We show the general evolution of the radiation properties using model \texttt{m7a0-M22} in \autoref{fig:heatmaps_Trad}. When the disk is geometrically thick, the outermost photosphere radius on the order of $\sim 100-300r_g$, with a radiation temperature of $T_{\rm rad}\sim 10^{4.75}$ K (UV) at the outer edge and $T_{\rm rad}\sim 10^{5.5}$ K (soft X-rays) in the corona/funnel region, respectively. The disk later vertically collapses, with the outer photosphere radius shrinking to $\sim 60 r_g$ as the disk radially contracts. At this stage, the coldest radiation from the outer photosphere edge still peaks in the UV ($T_{\rm rad}\sim 10^{4.75}$ K), but the radiation originating from the corona near the BH cools substantially, reaching roughly $T_{\rm rad}\sim 10^{5}$ K. This drastic and rapid change in radiation temperature from the corona during collapse is consistent with state transitions in the X-ray. The radiation/spectral properties are discussed in detail in \autoref{sec:spectra}.

\subsection{Thermal Instability and Disk Collapse}

The thin disk solution is thermally unstable due to the difference in temperature dependence of the heating and cooling functions \citep{1995ApJ...438L..37A}. Depending on the properties of the inflow, magnetic field, and outflow, disks may become thermally unstable as the accretion rate approaches near Eddington values \citep{2011ApJ...732...52Z,2014ApJ...786....6L,2019ApJ...887..256H,2022ApJ...930..108W,2023ApJ...954..150H}. To date, several numerical studies have attempted to study transitions of thick disks from either low-luminosity or super-Eddington accretion disks into a thin disk under varying magnetic field strengths and configurations \citep{2015MNRAS.447...49S,2021ApJ...919L..20D,Curd2023,2024ApJ...966...47L}. One major accomplishment of simulations has been the determination that magnetic pressure support is possible when strong magnetic fields thread the disk, such as in a magnetically arrested disk \citep[MAD,][]{Narayan2003}. In SANE disks, the fields are weaker and no such pressure support is expected.

To better understand the energy transport in each simulation, we first attempt to quantify the advection of radiation versus vertical diffusion in a similar manner to \cite{2016MNRAS.456.3929S}. We compute the radial and angular fluxes at the edges of a box with $r_1=15r_g$, $r_2=20 r_g$, $\vartheta_{1,2}=\pi/2 \pm 0.05$. Defining the luminosity at the radial edges as
\begin{equation}
    L_{r,(1,2)}=2\pi \int_{\vartheta_1}^{\vartheta_2}R^r_t \sqrt{-g} d\theta 
\end{equation}
and the angular edges as
\begin{equation}
    L_{\vartheta}=4\pi \int_{r_1}^{r_2}R^\vartheta_t \sqrt{-g}dr,
\end{equation}
where we have integrated both the top and bottom surfaces for $L_\theta$ assuming symmetry about the mid-plane. The difference $L_{r,1}-L_{r,2}$ estimates the amount of radiation advected from $r_2$ to $r_1$. The total radiation energy generated within the box is $L_{r,1}-L_{r,2} + L_\vartheta$. We follow \cite{2016MNRAS.456.3929S} and define the advection coefficient 
\begin{equation}
    q_{\rm adv}=\dfrac{L_{r,1}-L_{r,2}}{L_{r,1}-L_{r,2} + L_\theta}.
\end{equation}
Estimating this quantity for each model between $\Delta t=5.7-8.5$ days, we find $0.08 < q_{\rm adv} < 0.17$. Therefore, radiation is not efficiently trapped by the inflowing gas and instead is able to diffuse out vertically.

Indeed, this can also be seen by estimating the accretion timescale for gas versus the diffusion time for radiation. The accretion timescale at a given radius may be roughly approximated as 
\begin{equation}
    t_{\rm acc}=-\dfrac{R}{u^r},
\end{equation}
where $u^r$ is the radial velocity of the inflow. The diffusion timescale, or the time it takes for a photon generated at the disk mid-plane to escape vertically through the disk, is given by
\begin{equation}
    t_{\rm diff}\approx 3H\tau_{\rm tot},
\end{equation}
where $H$ is the disk height and $\tau_{\rm tot}$ is the total optical depth through the disk from $\vartheta=0$ to $\vartheta=\pi/2$. For each model, we estimate each timescale between $\Delta t=5.7-8.5$ days. We find that over the bulk of the disk ($r>r_{\rm ISCO}$), the ratio between the two is $1.5 \lesssim t_{\rm diff}/t_{\rm acc} \lesssim 4$. Therefore, radiation is expected to be able to diffuse during accretion instead of being efficiently trapped. For comparison, model \texttt{300a0} in \cite{2015MNRAS.447...49S} has $\dot{M}_\bullet \approx 9.6\dot{M}_{\rm Edd}$ and $t_{\rm diff}/t_{\rm acc}\gtrsim10$.

A precise determination of the stability and energetics is beyond the scope of this study. However, it is conceivable that the outflows, which are stronger as BH spin increases from $-0.9$ to $0.9$ (see \autoref{fig:scalars}), prevent or delay the disk becoming thermally unstable \citep{2019ApJ...887..256H,2023ApJ...954..150H}. The effects that BH spin has on disk stability in systems with winds and magnetic fields should be examined in further detail, as our simulations hint at higher spin, SANE systems being more stable.

\subsection{Time Evolution of Thermal Emission Properties} \label{sec:spectra}

\begin{figure}
    \centering{}
    \includegraphics[width=\columnwidth]{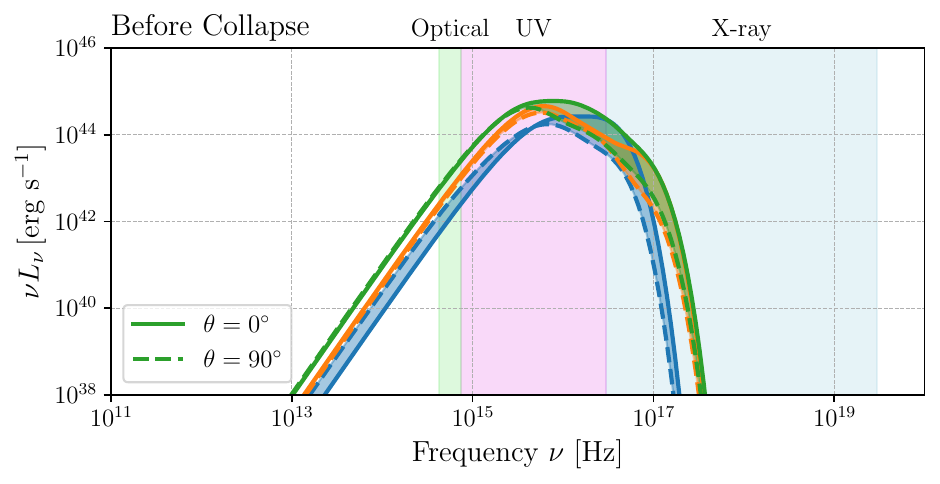}\\
    \includegraphics[width=\columnwidth]{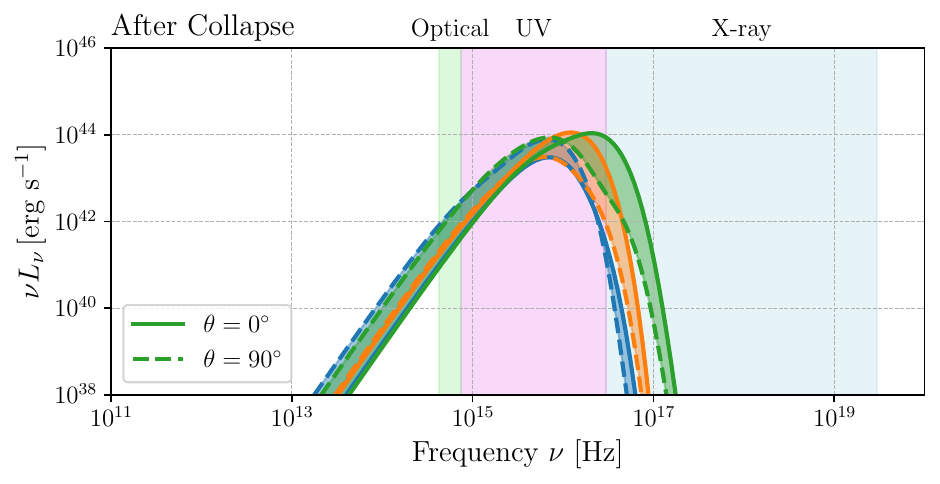}
    \caption{Here we show the spectrum for each BH spin before disk collapse ($\Delta t=5.7$ days, top) and after disk collapse ($\Delta t=\Delta t_{\rm final}$, bottom). Note $\theta=0^\circ$ (viewer at +z) is shown as the solid lines while $\theta=90^\circ$ (viewer at +x) is shown as the dashed lines. Viewing angle effects lead to obscuration of X-ray photons when the disk is viewed edge-on ($\theta = 90^\circ$). Note that the $a_\bullet=-0.9$ model has the weakest soft X-ray emission while the spin $a_\bullet=0.9$ model is brighter due to ISCO effects.}
    \label{fig:spectrum_spins}
\end{figure}

\begin{figure}
    \centering{}
    \includegraphics[width=\columnwidth]{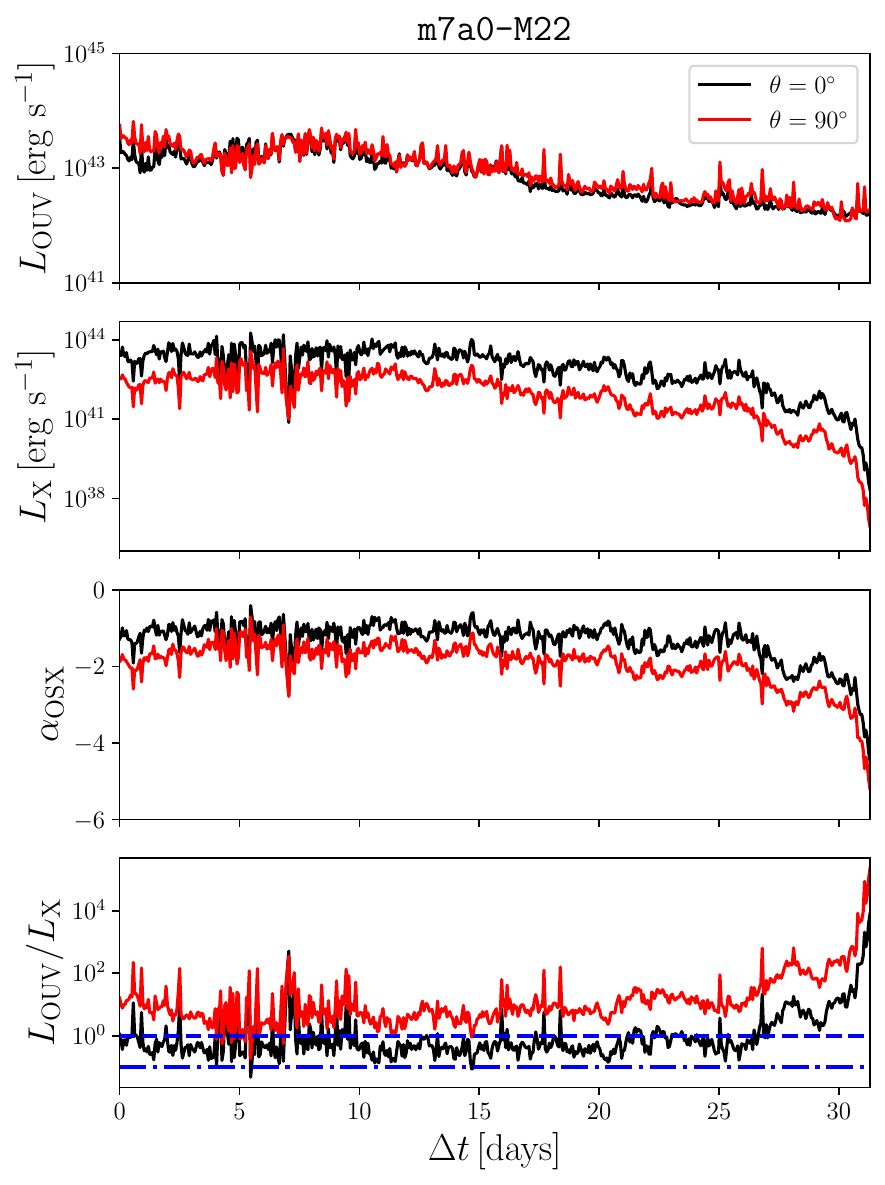}
    \caption{Here we show the spectral properties of model \texttt{m7a0-M22}. We indicate $L_{\rm OUV}/L_{\rm X}=1$ (dashed blue line) and $L_{\rm OUV}/L_{\rm X}=0.1$ (dash-dotted blue line) in the bottom panel for ease of comparison.}
    \label{fig:Lvst_a0}
\end{figure}

\begin{figure}
    \centering{}
    \includegraphics[width=\columnwidth]{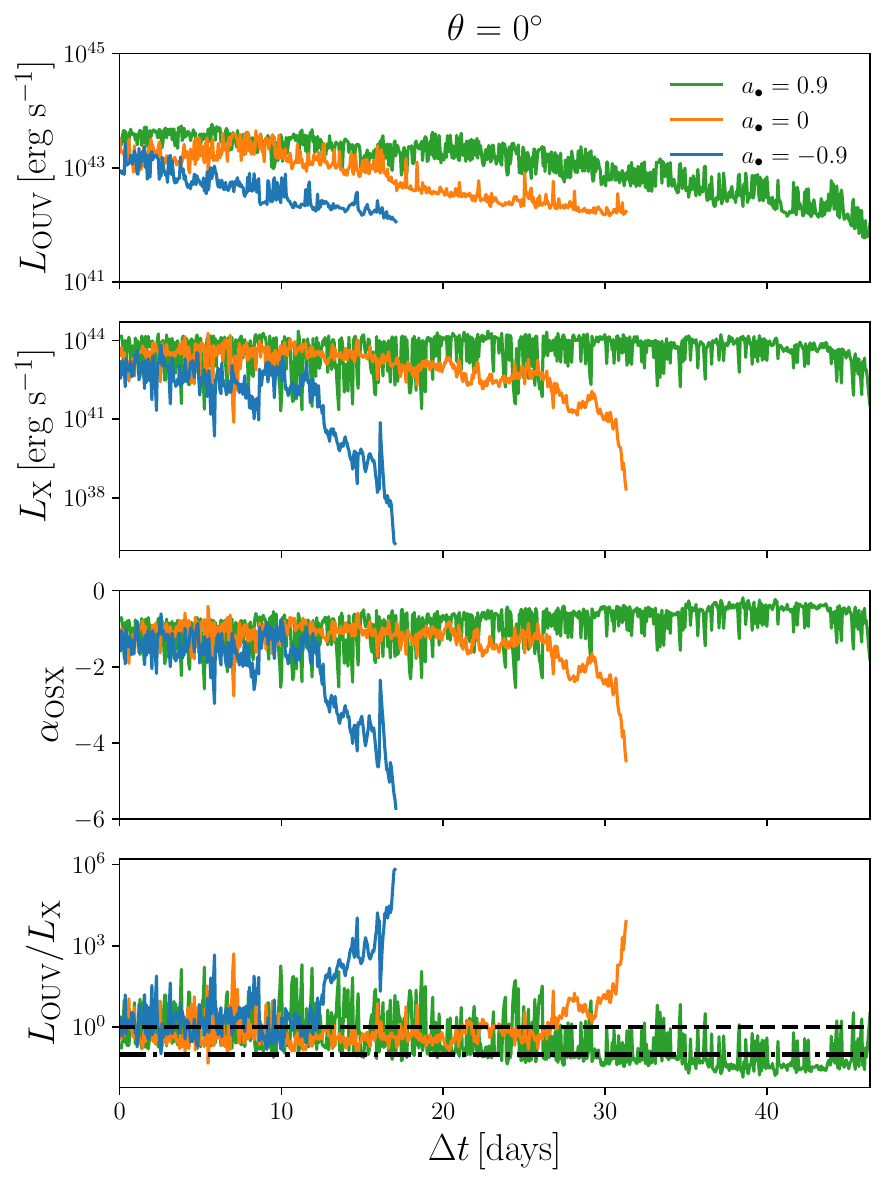}
    \caption{Here we show the spectral properties for all models for an observer at $\theta=0^\circ$. The $a_\bullet=0.9$ model shows the brightest soft X-ray emission, but all models have a similar value for $\alpha_{\rm OSX}$ prior to disk collapse. All models show rapid OUV and X-ray variability prior to collapse. We indicate $L_{\rm OUV}/L_{\rm X}=1$ (dashed black line) and $L_{\rm OUV}/L_{\rm X}=0.1$ (dash-dotted black line) in the bottom panel for ease of comparison.}
    \label{fig:Lvst_all_th0}
\end{figure}

\begin{figure}
    \centering{}
    \includegraphics[width=\columnwidth]{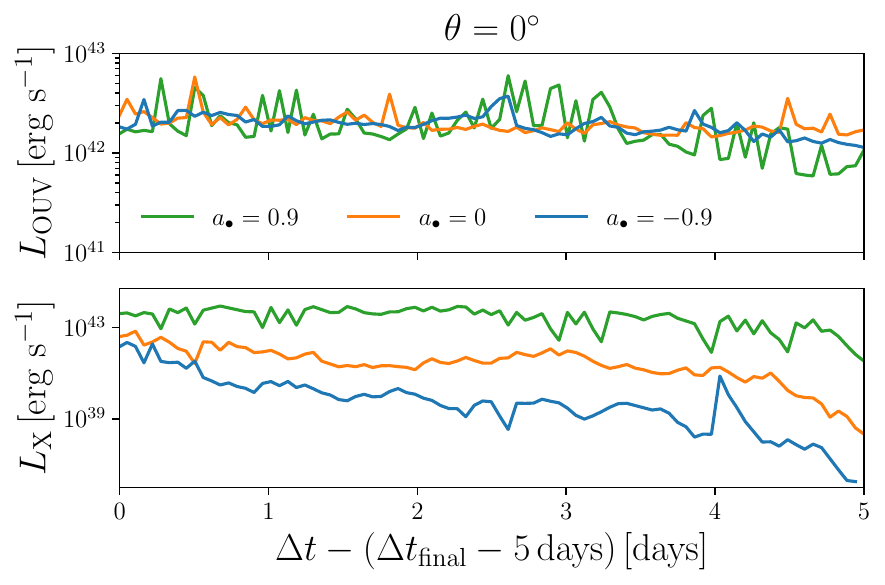}
    \caption{Here we zoom in on the band luminosities for all models in the last 5 days for an observer at $\theta=0^\circ$.}
    \label{fig:Lvst_all_th0_zoom}
\end{figure}

The version of the \textsc{koral} code used in this work computes frequency integrated (or monochromatic) radiative quantities. It also assumes thermal energy distributions of the ions and electrons when calculating absorption and emission coefficients. This assumption can be inaccurate in the jet region where $\sigma>1$ and electrons can be accelerated into a non-thermal distribution \citep[e.g.][]{2017MNRAS.470.2367C}. Assuming thermal radiation in the optically thick disk/wind region is more accurate, so we apply a simple viewing angle dependent thermal emission model to the \textsc{koral} simulation data. In \cite{Curd2019}, we adopted the \textsc{heroic} post-processing code to solve for a more accurate equilibrium radiation solution; however, the \textsc{heroic} code is computationally expensive for even a single snapshot of GRRMHD data. For this work, we are interested in calculating the spectrum of every snapshot in a dataset of 1,665 snapshots, making the use of such an expensive code impractical.

For thermal emission, we assume the radiation originates from the photosphere calculated using the electron scattering opacity (see \autoref{eq:taures} and \autoref{eq:tauthetaes}). At the mild accretion rates used in this work, there are viewing angles where no radial photosphere surface exists (see \autoref{fig:heatmaps_Trad}). For these viewing angles, we adopt the angle integrated $\tau_\vartheta=1$ surface as the emission surface. This step is crucial since the soft X-rays are emitted from the hot inner accretion disk.

Although the \textsc{koral} code uses a frequency integrated radiation solver, important effects such as Comptonization are captured and modify the total radiation energy density $\widehat{E}$. Assuming the photons making up the radiation field are in a thermal distribution, the radiation temperature can be computed directly from the \textsc{koral} data using the radiation energy density (\autoref{eq:Trad}). We use the frequency integrated luminosity exiting the emission surface ($dL_r$ or $dL_\vartheta$) directly from the \textsc{koral} data to normalize the distribution. 

Since it is possible for observers viewing the system from near edge on angles to receive emission from parts of the disk where the funnel boundary faces them, it is necessary to perform the emission calculation in 3D. We do this by simply assuming the disk is symmetrical in $\varphi$, which simply means we copy the 2D $(r,\vartheta)$ data onto a 3D $(r,\vartheta,\varphi)$ grid with all fluid and radiation variables held constant in $\varphi$, and account for viewing angle effects. Viewing angle effects enter once we assume a distant observer whose camera sits at polar coordinates $(\theta,\phi)$. Note that these coordinates are distinct from the fluid coordinates $(r,\vartheta,\varphi)$. Since the disk is symmetric and aligned with the BH spin, we simply use $\phi=0$. Assuming blackbody emission, the frequency dependent luminosity from each cell is then
\begin{equation}
    dL_{r,\nu}(\nu)=dL_r f_r(\vartheta,\varphi,\theta,\phi)\dfrac{15}{\pi^4 \nu_p}\left(\dfrac{\nu}{\nu_p}\right)^3\left(e^{\nu/\nu_p}-1\right)^{-1}
\end{equation}
for cell faces where $\boldsymbol{\widehat{r}}$ is the surface normal and 
\begin{equation}
    dL_{\vartheta,\nu}(\nu)=dL_\vartheta f_\vartheta(\vartheta,\varphi,\theta,\phi)\dfrac{15}{\pi^4 \nu_p}\left(\dfrac{\nu}{\nu_p}\right)^3\left(e^{\nu/\nu_p}-1\right)^{-1}
\end{equation}
for cell faces where $\pm \boldsymbol{\widehat{\vartheta}}$ is the surface normal. Here $\nu_p=kT_{\rm rad}/h$, and $f_r(\vartheta,\varphi,\theta,\phi)$ and $f_\vartheta(\vartheta,\varphi,\theta,\phi)$ account for viewing angle effects. The radial term is given by
\begin{equation}
 f_r(\vartheta,\varphi,\theta,\phi=0) = \cos \vartheta \cos \theta + \sin \vartheta \sin \theta \cos \varphi,
\end{equation}
and the angular term is given by
\begin{equation}
 f_\vartheta(\vartheta,\varphi,\theta,\phi=0) = \cos \vartheta \cos \theta + \sin \vartheta \sin \theta \cos(\varphi - \pi/2).
\end{equation}
Note that we have explicitly set $\phi=0$, which simplifies the result slightly. To account for the obscuration of rays from the disk edge that must traverse through the disk material to reach the observer (for example, the disk edge at $\varphi=\pi$), we set $f_r=0$ or $f_\vartheta=0$ if the ray intersects the disk on its way to the observer or if the term becomes negative.

To compute spectra, we sum up the emission from each cell over the entire $\tau=1$ surface. We show example spectra from each model to demonstrate viewing angle effects as well as how spin modifies the spectrum before and after disk collapse in \autoref{fig:spectrum_spins}. Prior to disk collapse, models show a soft X-ray excess and a peak in the far UV, which is in agreement with the slim disk model for above Eddington accretion rate disks \citep[e.g.][]{1999ApJ...522..839W}. Viewing angle effects lead to some obscuration, but even our simple thermal radiation model predicts significant soft X-rays when $\theta=90^\circ$. Similar to \cite{Curd2019}, this excess occurs due to a hot, optically thick wind. After disk collapse, the soft X-rays diminish significantly and the spectrum resembles the modified blackbody in the classical thin accretion disk solution, peaking in the far UV. Note that the peak frequency moves from the far UV towards the X-ray as BH spin increases, in agreement with ISCO effects in the thin disk model.

To compare our simulated spectra with observed TDEs such as AT2021ehb, we compute band integrated luminosities $L_{\rm OUV}$ and $L_{\rm X}$. For $L_{\rm OUV}$, we integrate over the range $4.87\times10^{14}$ Hz to $1.56\times10^{15}$ Hz. For $L_{X}$, we integrate over photon energies in the range 0.03 keV ($7.25\times10^{16}$ Hz) to 10 keV ($2.42\times10^{18}$ Hz). We also compare the luminosity between the optical/UV and soft X-ray via the quantity \citep[][Equation 8]{Gezari2021}
\begin{equation} \label{eq:alphaoOSX}
 \alpha_{\rm OSX} = \dfrac{\log[L_\nu(2500 {\text{\AA}})/L_\nu(0.5 {\rm{keV}})]}{\log[(\nu(2500 {\text{\AA}})/\nu(0.5 {\rm{keV}})]}.
\end{equation}
This gives the slope of the flux density $f_\nu \propto \nu^{\alpha_{\rm OSX}}$ between $\nu=2500 {\text{\AA}}$ and $\nu=0.5$ keV and is useful for comparison with observations. 

We present spectral parameters $L_{\rm OUV}$, $L_{\rm X}$, and $\alpha_{\rm OSX}$ as a function of time for model \texttt{m7a0-M22} in \autoref{fig:Lvst_a0}. We also compare these parameters for all values of BH spin considered for an observer at $\theta=0^\circ$ in \autoref{fig:Lvst_all_th0}. The optical/UV luminosity is not much effected by the viewing angle, which is expected since the optical/UV component is mostly emitted by the larger radius disk edge. The soft X-rays on the other hand see a nearly order of magnitude decrease in brightness when the disk is viewed edge-on ($\theta=90^\circ$). Despite our rather crude spectral calculation, we find peak values for $\alpha_{\rm OSX}$ in the range of $\sim -2$ to $-1$ and this remains true even for BH spin $a_\bullet=\pm0.9$ (\autoref{fig:Lvst_all_th0}). This is similar to that of AT2021ehb prior to collapse. It is also typical of TDEs with detectable soft X-rays \citep{Gezari2021}. \textcolor{black}{An important caveat of our thermal spectra is that the $\tau=1$ surface only captures thermal emission from the disk and optically thick wind components. While X-rays are expected to decline in general upon collapse (e.g. consider the radiation temperature in the funnel in \autoref{fig:heatmaps_Trad}), our values for $L_{\rm X}$ and $\alpha_{\rm OSX}$ become inaccurate once the disk becomes thin. Comptonization of disk photons and X-ray reflection are expected to contribute significantly to X-rays in thin disks. This is especially true in models where $|a_\bullet|>0$ \citep[e.g.][]{2025ApJ...981..144R}. In short, the decrease in X-ray luminosity and increase in $\alpha_{\rm OSX}$ that our spectra depict is likely overestimated and should only be taken as a qualitative feature. More explicitly, the observed decline in $L_{\rm X}$ in \autoref{fig:Lvst_all_th0} should not be interpreted as the disappearance of an inverse comptonizing corona, but rather a decrease in the emission from the corona. The optical and UV emission is more accurate since it is produced by the disk edge, which is adequately described by thermal emission.}

The BH spin modifies the spectral properties in several ways. Firstly, the X-ray luminosity prior to collapse is generally higher as BH spin increases from retrograde ($a_\bullet=-0.9$) to prograde ($a_\bullet=0.9$). Second, the $a_\bullet=0.9$ model appears to have rapid and intense variability in comparison to both the retrograde and non-spinning models. This is possibly due to variability on the horizon scale; however, a full 3D calculation is necessary to precisely measure variability. 2D models can overestimate variability since gas may cool or become obscured at one patch of $\varphi$, but hot unobscured gas at another $\varphi$ could still be emitting significant X-rays. Second, the X-ray decay appears to become more sudden as BH spin increases. The $a_\bullet=-0.9$ and $a_\bullet=0$ models show a two stage decay where the X-rays first decrease by roughly an order of magnitude, and then either steadily decay further ($a_\bullet=-0.9)$ or stabilize for a few days before the disk completely collapses $(a_\bullet=0$). This effect is likely due to disk annuli at larger radius collapsing simultaneously as spin increases (\autoref{fig:heatmaps_ISCOs}). Lastly, it can also be seen when zooming in to the last 5 days of each simulation (\autoref{fig:Lvst_all_th0_zoom}) that the optical/UV in the $a_\bullet=0.9$ model appears to have a factor of a few decrease in the final day, whereas the other models remain nearly constant even during the start of the X-ray decline and after collapse in the final day.

\begin{figure}
    \centering{}
    \includegraphics[width=\columnwidth]{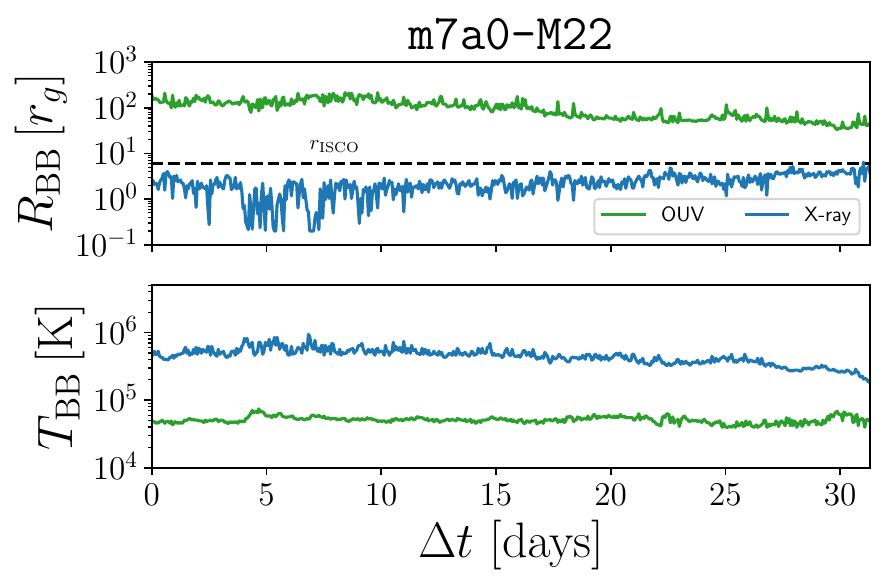}\\
    \includegraphics[width=\columnwidth]{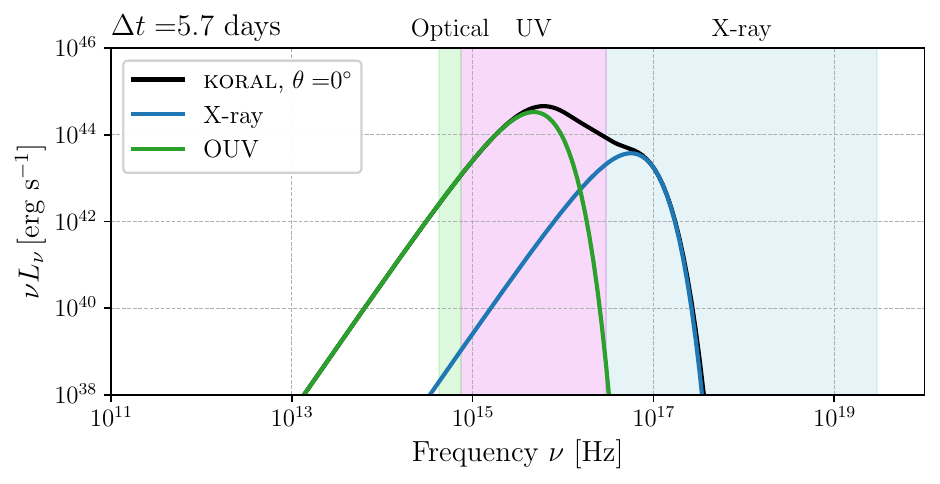}    
    \caption{Here we show the blackbody fits for radius (top) and temperature (middle) in the optical/UV band (green) and soft X-ray band (blue). The viewing angle is $\theta=0^\circ$. We show the range of values obtained with our luminosity constraints for the optical/UV fits with the green shaded region. The X-ray blackbody radius appears to approach the ISCO radius (black dashed line) as the disk collapses. In the bottom panel we compare the blackbody fits with the input \textsc{koral} spectrum at $\Delta t = 5.7$ days.}
    \label{fig:BBfits_a0}
\end{figure}

\begin{figure}
    \centering{}
    \includegraphics[width=\columnwidth]{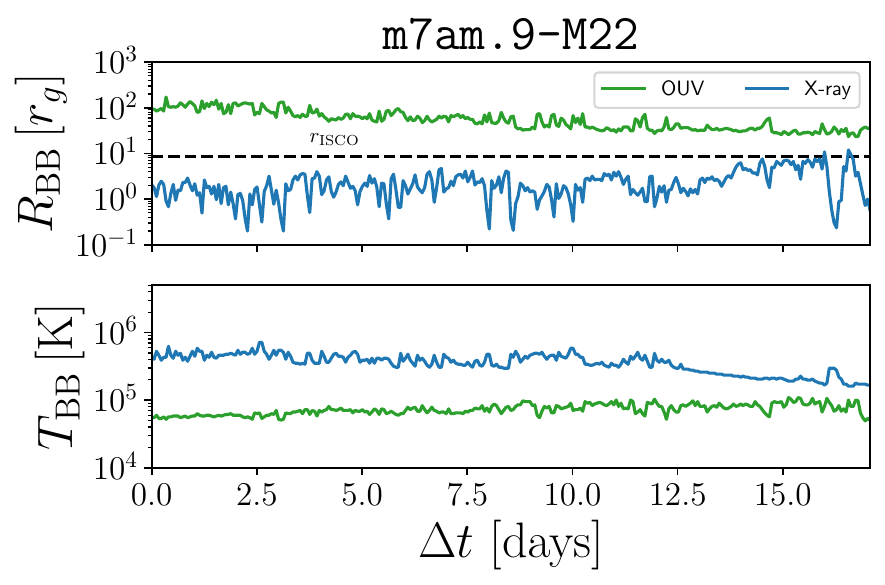}\\
    \includegraphics[width=\columnwidth]{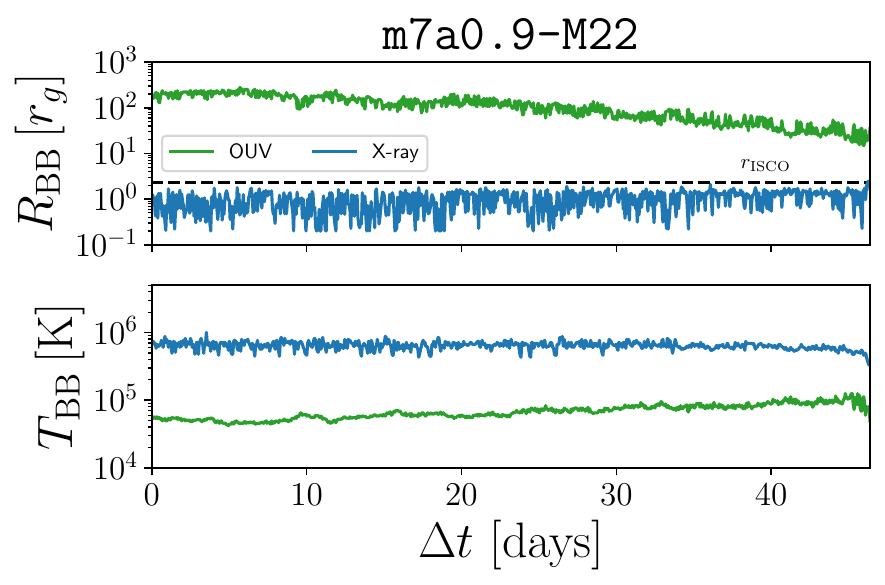}    
    \caption{Here we show the blackbody fits for $a_\bullet=-0.9$ (top) and $a_\bullet=0.9$ (bottom). The viewing angle is $\theta=0^\circ$.}
    \label{fig:BBfits_other}
\end{figure}

It is interesting to consider how blackbody fits in the observable bands from our model spectra compare with typical values seen in the literature. As such, we attempt to fit a blackbody spectrum to both the optical/UV band ($4.87\times10^{14}$ Hz to $1.56\times10^{15}$ Hz) and the X-ray band (0.03 keV to 10 keV) separately. We fit each individual \textsc{koral} spectrum at time $\Delta t$ at viewing angles $\theta=0, 30, 60,$ and 90 degrees. For each band, we consider a spherical, optically thick sphere of gas with radius $R_{\rm BB}$ and temperature $T_{\rm BB}$. We then perform a $\chi^2$ minimization by varying $R_{\rm BB}$ and $T_{\rm BB}$ and comparing the Planck spectrum in the respective band versus our post-processed \textsc{koral} spectrum (as shown in \autoref{fig:spectrum_spins}). In the X-ray band, the break where the Wien tail of the spectrum begins places strong constraints on the blackbody fitting procedure. For the optical/UV bands, the peak of our frequency dependent spectra are in the far UV and no break is available, which can lead to degeneracy in the $(R_{\rm BB},T_{\rm BB})$ parameter space. As such, the best we can do is use reasonable limitations for the peak luminosity of the blackbody used to fit the Rayleigh Jean's tail. For simplicity, we assert that the model can only have a peak luminosity of the blackbody model $L_{\rm BB}(\nu=2.821kT_{\rm BB}/h)$ that is between $2L(\nu=1.56\times10^{15}$ Hz) and $L_{\rm Edd}$. Note $L=\nu L_\nu$ from the post-processed \textsc{koral} spectrum (as shown in \autoref{fig:spectrum_spins}, for example).

The results of our fitting procedure are shown for model \texttt{m7a0-M22} in \autoref{fig:BBfits_a0}. The time evolution of the extracted radius and temperature illustrate that X-ray fluctuations arise due to rapid variation in the photosphere scale and temperature. Furthermore, the temperature of the optical/UV photosphere is nearly constant. In fact, the temperature is more representative of the \textsc{koral} data as the disk begins to collapse since the luminosity drops below Eddington, and this value is nearly constant at $T\sim 4\times 10^4$ K. 

The photosphere can be described as follows. The optical/UV photons originate from a $r\sim 60-100 r_g$ surface that radially shrinks at a nearly constant temperature of $T\approx 4\times 10^4$ K. These radii are consistent with the photosphere radii shown in \autoref{fig:heatmaps_Trad}. Similar to \cite{Thomsen2022}, we find that $R_{\rm BB}$ and $T_{\rm BB}$ for the optical/UV is insensitive to viewing angle. The X-ray photons originate from plasma near the BH ($r\sim 0.1-1r_g$) that is much hotter ($T\approx 5-7\times10^5$ K) and cools rapidly upon collapse to $T\approx 2-3\times 10^5$ K. Precise values depend on the spin of the BH (see \autoref{fig:BBfits_other}). Perhaps the most striking result is the late X-ray blackbody radius. For all values of $a_\bullet$, the blackbody radius in the X-rays approaches the ISCO radius upon collapse.

Another interesting observation in our simulations is why the blackbody radius is occasionally less than $\sim 1r_g$ at various points. The periods in time where $R_{\rm X, BB}\sim 0.1$ occurs when the optical depth along lines of sight that see the inner disk increases due to variable winds at the disk-wind interface. Interestingly, no such decrease in $T_{\rm X, BB}$ coincides with this phenomenon. As such, all of the X-rays that appear in the spectrum are X-rays which have been partially absorbed through the wind before reaching the observer. 

\begin{figure*}
    \centering{}
    \includegraphics[width=\textwidth]{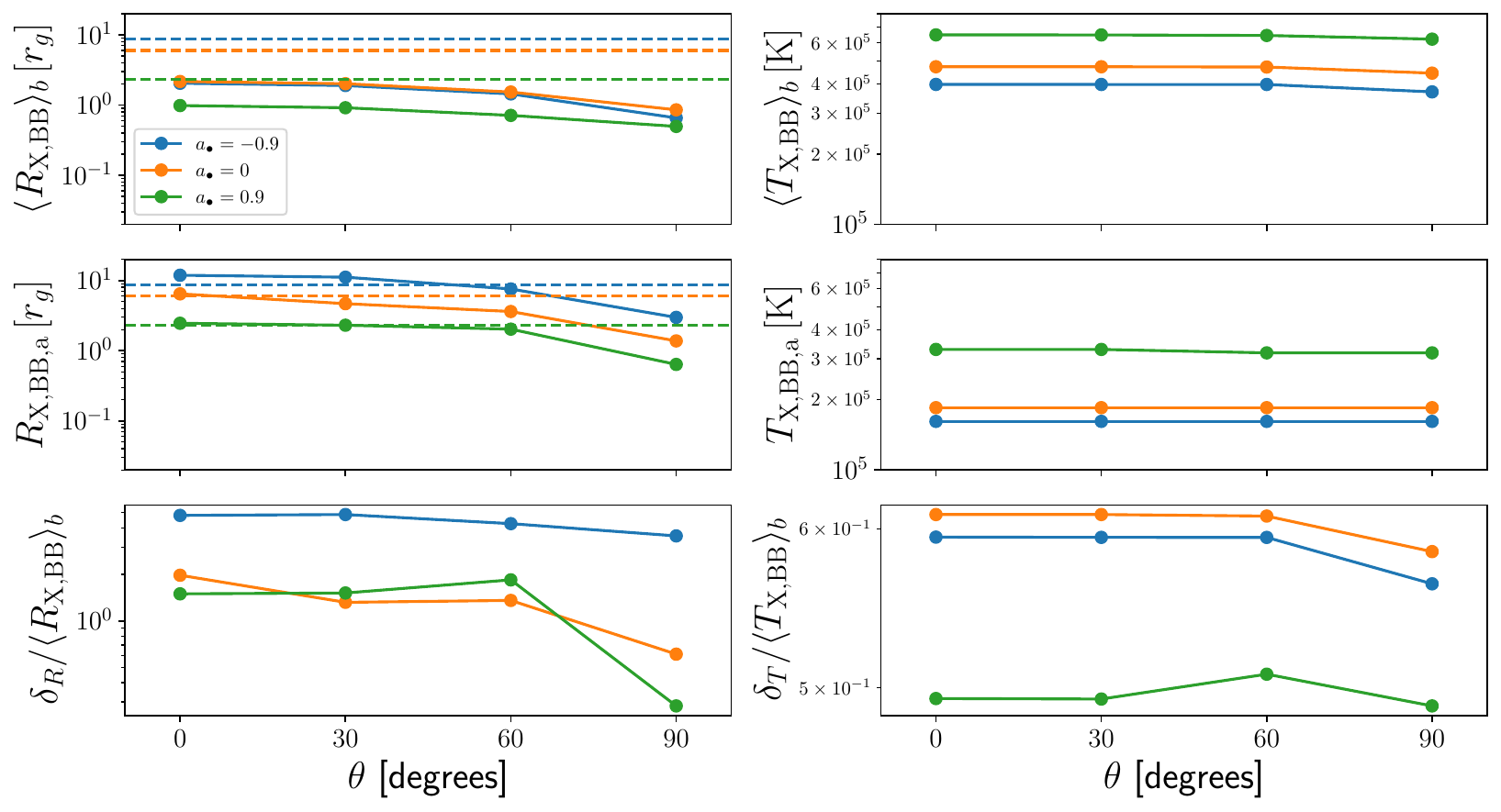}  
    \caption{Here we show the properties of the blackbody fits for the soft X-ray as a function of viewing angle. The ISCO radius is indicated with a dashed horizontal line for each value of BH spin. Note that subscript '$b$' and subscript '$a$' define quantities as \textit{before} and \textit{after} disk collapse. More precisely, quantities with subscript '$b$' are averaged over $0< \Delta  t\leq\Delta t_{\rm final}-5$ days and subscript '$a$' are the maximum radius (minimum temperature) for the time range $\Delta t_{\rm final}-5 \, {\rm days} < \Delta t \leq \Delta t_{\rm final}$.}
    \label{fig:BBfits_stats}
\end{figure*}

To illustrate the effects of BH spin and viewing angle on the black body fits in the X-rays, we provide metrics of $R_{\rm X,BB}$ (and $T_{\rm X,BB}$) and the change in $R_{\rm BB}$ before (subscript '$b$') and after (subscript '$a$') collapse. We define $\langle R_{\rm X,BB} \rangle_b$ as $R_{\rm X,BB}$ averaged over $0< \Delta  t\leq\Delta t_{\rm final}-5$ days and $R_{\rm X,BB,a}$ as the maximum value of $R_{\rm X,BB}$ for the time range $\Delta t_{\rm final}-5 \, {\rm days} < \Delta t \leq \Delta t_{\rm final}$. We show the difference in $R_{\rm X,BB}$ before and after collapse ($\delta_R\equiv |R_{\rm X,BB,a}-\langle R_{\rm X,BB} \rangle_b|$) in \autoref{fig:BBfits_stats}. We also perform a similar calculation on the temperature except we take the minimum temperature to be $T_{\rm X,BB,a}$. The reasoning behind our choice for $R_{\rm X,BB,a}$ and $T_{\rm X,BB,a}$ is that we must stop the simulation once the disk becomes too thin, but in reality we expect the disk to reach a new equilibrium described approximately by our collapsed state beyond when we stop the simulation. We have normalized the relative differences by quantities before the collapse ($\langle R_{\rm X,BB} \rangle_b$ and $\langle T_{\rm X,BB} \rangle_b$) to obtain unitless quantities. 

Our results show a correlation with $R_{\rm X,BB}$, $T_{\rm X,BB}$, $\delta_R$, $\delta_T$, and the BH spin. Interestingly, we obtain a different trend than \cite{Thomsen2022}, who found that blackbody fits give a slightly increasing radius as viewing angle increases. \cite{2025MNRAS.539.3473Q} find that the radius decreases  with viewing angle; however, they obtain substantially larger blackbody radii ($R_{\rm X,BB}\sim10r_g$) than ours. The discrepancies between our work and \cite{Thomsen2022,2025MNRAS.539.3473Q} is potentially due to the fact that their accretion rates were much higher. Additionally, \cite{2025MNRAS.539.3473Q} perform their simulations in 2D without magnetic fields. We also note that we use a simple blackbody approach as opposed to Monte Carlo radiative transfer.

The relative differences $\delta_R$ and $\delta_T$ are fairly insensitive to viewing angle, with the exception of $\theta=90^\circ$, which suggests information on the BH spin can be imprinted on the spectral features regardless of viewing angle once the X-rays become detectable. Future work is needed to determine how these quantities evolve with BH spin as we only simulate three values of $a_\bullet$. Nevertheless, our results suggest that observations with well constrained measurement of the BH mass and viewing angle combined with finer time resolution, which would improve accuracy in determining $R_{\rm X,BB}$ during the disk collapse phase, may reveal information about the BH spin.

\section{Discussion}

Accretion disk state transitions have long been expected in TDEs since the fallback rate eventually becomes sub-Eddington. Accretion theory can complicate this simple picture depending on the magnetic field properties \citep[e.g.][]{Lancova2019,2022ApJ...935L...1L,Curd2023}. In addition, the fact that AT2021ehb is currently the only non-jetted TDE to display a state transition suggest that magnetic pressure support may be taking place in many TDEs \citep{2025ApJ...993...14A}. Alternatively, state transitions may be commonplace, but occurring on short timescales undetectable by current campaigns \citep[e.g.][]{2025ApJ...985...77P}. 

In this work, we have studied state transitions of SANE accretion disks, which have similar properties to many thermal TDEs \citep{Curd2019}. Our disks are formed from initial conditions similar to the late-stage CEM presented in \cite{Metzger2022}. The thermal optical/UV and soft X-ray behavior observed in our model spectra are qualitatively consistent with many thermal TDEs in the literature.

Prior to collapse, $\alpha_{\rm OSX}$ ranges from roughly -1 to -2. We also observe that $\alpha_{\rm OSX}$ increases over time, which is consistent with ASASSN-14li, ASASSN-15oi, AT2018fyk/ASASSN-18ul, AT2019azh/ASASSN-19dj, AT2019ehz/Gaia19bt, and AT2021ehb \citep{Gezari2021,2022ApJ...937....8Y}. During the collapse, we observe a sharp decline in $\alpha_{\rm OSX}$. While observed TDEs also show a decline, it is much less dramatic. For example, AT2021ehb only showed a decrease in $\alpha_{\rm OSX}$ of $\sim 0.5$. We attribute this to two factors. First, the \textsc{koral} simulations assume a thermal electron distribution, which underestimates coronal X-rays. Second, our post-processed spectra only treat radiation from the $\tau=1$ surface and do not incorporate coronal emission or reflected X-rays.

Rapid X-ray variability appears in both $L_{\rm X}$ and $\alpha_{\rm OSX}$. We do not calculate the periodicity of the variability due to the simulations being carried out in 2D; however, our results confirm that X-ray variability can appear in the thermal radiation from the accretion disk. Our emission region analysis demonstrates that fluctuation in the scale and temperature of the emission zone may be responsible for X-ray variability. Ultimately, these phenomena connect to variability in the accretion rate itself (\autoref{fig:scalars_mdot}).

The short timescale evolution of the X-rays as the disk collapses may reveal information about the BH spin. AT2021ehb appears to have a two stage decline with different hardness ratios (HRs) at each stage \citep{2022ApJ...937....8Y}. The first decline starts at $\delta t = 271$ days with the X-ray luminosity first rapidly dropping from $\sim 7\times 10^{43} {\rm erg\, s^{-1}}$ at $\delta t = 271$ days to $\sim 4\times 10^{42} {\rm erg\, s^{-1}}$ at $\delta t = 273.7$ days. At the same time, the HR also rapidly declined from $\sim 0.5$ to $\sim -0.5$. This decline is likely due to Coronal changes and our models cannot be applied here. However, the second X-ray decline between $\delta t = 320.9$ days ($L_{\rm X} \sim 4\times 10^{42} {\rm erg \, s^{-1}}$) to $\delta t = 327.2$ days ($L_{\rm X} \sim 5\times 10^{41} {\rm erg \, s^{-1}}$) occurs at a nearly constant HR of $\sim -0.5$ and the X-rays appear to be approximately thermal. Our thermal only emission models can be more appropriately applied here.

In our simulations, the nature of the X-rays during disk collapse changes as the BH spin increases from $-0.9$ to $0.9$ due to larger radii of the disk collapsing at once. This leads to the X-rays having a steadier decline over the course of $\sim 5$ days in our $a_\bullet=-0.9$ and $a_\bullet=0$ models. However, the $a_\bullet=0.9$ model rapidly declined in its X-rays within a day. Unfortunately, the collapse time of $\sim 5$ days is shorter than the time between observations in AT2021ehb when it declined in the thermal X-rays stage (roughly 7 days). High cadence observations in future observing campaigns may be useful in determining the BH properties in TDEs.

Our thermal spectra give an inferred radius from the soft X-rays of $\sim r_g$ prior to collapse and $\sim r_{\rm ISCO}$ after collapse. The pre-collapse radius is consistent with a few thermal TDEs \citep{Gezari2021}. The radius and temperature inferred from the X-rays in AT2021ehb after its state transition at $\delta t = 327.2$ days are $R_{\rm in}= 0.5_{-0.2}^{+38.8} r_g$ and $T= 5.9_{-1.9}^{+3.9}\times 10^5$ K (Yuhan Yao, private communication). The range given is consistent with all radii in \autoref{fig:BBfits_stats}, which unfortunately makes estimating the BH spin impossible without more accurate radii. Additionally, our temperatures all fall outside of their estimated range. Our temperatures are likely colder since we omit coronal effects in our emission calculation. We also note that our pre-collapse radius of $\sim r_g$ is inconsistent with the pre-state transition radii in AT2021ehb, which are generally of the order $\sim10^{-1.4}r_g$. \cite{2022ApJ...937....8Y} note that Compton scattering or the emission structure itself may contribute to the small radii measured in AT2021ehb.\footnote{Our spectra when the optical depth towards the inner disk increases (when $R_{\rm X, BB}\sim 0.1 r_g$) agree with this hypothesis; however, the average state of the accretion flow has low enough optical depth to give $\langle R_{\rm X,BB}\rangle\sim r_g$.} Nevertheless, the increase in X-ray emission radius as the disk collapses is consistent with the qualitative evolution of AT2021ehb, which suggests that a disk state transition is responsible for the X-ray decline. Emission models of accreting BHs that have persistent Compton thick scattering of X-rays (e.g. tilted disks or eccentric disks) may be more appropriate for studying the X-ray emission in TDEs where $R_{\rm X,BB} < r_g$, such as AT2021ehb. 


All of our model spectra peak in the far UV, well outside of observable ranges. Observations in the optical, near UV, and soft X-rays only will vastly underestimate the emission from such disks, even at the late time of $t>250$ days that we use for our initial conditions. These results agree with previous simulations \citep{Dai2018,Curd2019,Thomsen2022,2025MNRAS.539.3473Q}, but these are the first TDE disk simulations and spectra of $M_\bullet = 10^7$ BHs. The `missing-energy' problem associated with TDEs is easily explained if the large scale envelope emits mostly far UV photons.

\textcolor{black}{To-date, about 10\% of TDEs have been detected in the radio. Outflows from super-Eddington accretion flows \citep[$v\sim0.1-0.3c$,][]{Sadowski+2015b,Curd2023b} are faster than estimated outflow speeds for most thermal TDEs with early radio detections \citep[$v\sim0.008-0.07c$,][]{2024ApJ...971..185C}. For these slower winds that are promptly launched, it is likely that the stream self-intersection outflow powers the radio emission as it collides with the circumnuclear medium \citep{2020MNRAS.492..686L,2025ApJ...988L..24H}. For our simulations, the density weighted outflow speed (only considering the unbound wind with $\sigma<1$) prior to collapse is $v\sim0.1c$. This is consistent with several recently discovered delayed radio TDEs \citep{2024ApJ...971..185C}. The origin of the outflow responsible for radio emission in delayed radio TDEs is still uncertain; however, \cite{2021NatAs...5..491H} argue that a state transition to an enhanced accretion rate can explain the delayed radio emission with high kinetic energy outflows in ASASSN-15oi, for example.}

\textcolor{black}{Since the jet/wind power estimated from the accretion rate and efficiency (\autoref{fig:scalars}) is $P=\eta\dot{M}_{\bullet}c^2$, this suggests a phase of 10-40 days (depending on BH spin) with powerful outflows followed by a roughly 1-2 order of magnitude decline in jet/wind power coincident with the disk collapse. Provided the outflow is not impeded by debris and can shock on the circumnuclear medium, we expect this would result in a radio brightening followed by a steady decline once the outflow shuts off. While outflow properties prior to the $t>250$ days stage that we simulate is uncertain, if the system is described by the CEM with a thin disk at $r<R_{\rm circ}$ we expect weak outflows which may not be powerful enough to produce detectable radio emission to be present at this stage. We expect that the outflow properties of our simulations apply to thermal TDEs after disk formation despite the fact that we assume a SANE disk with an artificial dynamo. Several authors have demonstrated that the amplification of magnetic fields during TDE disk formation drives $\beta_{\rm mag}$ to values of 100 and below \citep{Sadowski2016,Curd2021,2026ApJ..1001...71A}.}

Finally, it is worth noting that in our simulations, the heating and mass contributed to the system by the returning debris stream is not captured since we initialize with a hydrostatic torus of gas. Recent work by \cite{2025ApJ...989...27T} suggests that the stream has an important effect on the photosphere radius and the mass accretion rate even at late times. In short, the collapse times that we find in this work may be delayed if the mass accretion rate remains elevated due to the stream. 

\section{Conclusions}

We find that all of our models are thermally unstable and diffusion of radiation vertically through the disk is quite efficient. Increasing BH spin delays the collapse by a similar time interval of $\sim14-15$ days as we increase the BH spin by $+0.9$ in \autoref{tab1}. This suggests that there is some fundamental property related to the disk stability that scales with BH spin, either through winds or perhaps the magnetic fields. We note; however, that in this study we have fixed the disk mass, not the accretion rate. Ergo, the mass accretion rate varies somewhat as BH spin changes.

Our models suggests that state transitions are expected in the CEM at late stages when the debris cloud becomes better described by a disk structure. The CEM could then explain the early- and late-time behavior of TDEs such as AT2021ehb provided that thermal instability of the disk is accounted for. Focusing on the near-Eddington pre-collapse stage, we also find general qualitative agreement with observations of TDEs with soft X-rays as (i) our models have a similar spectral slope, (ii) rapid X-ray variability appears in each model, and (iii) the scales of the optical/UV and soft X-ray emission regions when fit with a single blackbody are vastly different, with soft X-rays originating from $\sim r_g$ and the optical/UV originating from $\sim100 r_g$.

The analysis of our spectra and blackbody fitting tentatively suggests that constraining the BH spin may be possible by measuring the radius and temperature of the X-ray emitting region before and after state transitions since the inferred radius approaches $r_{\rm ISCO}$. However, a more precise study of the effects of BH spin on X-ray emission modeling requires a full 3D simulation, non-thermal electrons, and a full radiative transfer calculation. We employed 2D \textsc{koral} simulations and post-processed the data using a simple thermal emission model to reduce the computational cost. We plan to revisit the problem of state transitions in SANE disks with the aforementioned improvements in a future work.

\section{Acknowledgements}
 
We thank Brian Metzger for useful conversations as well as providing data tables which were used to establish our initial conditions. \textcolor{black}{We also thank the anonymous referee who provided insightful feedback that significantly improved the manuscript.} This work was supported by a grant from the Simons Foundation (00001470, BC, RA). 
Computational support was provided via ACCESS resources (grant PHY230006).

\section{Data Availability}
The data underlying this article will be shared on reasonable request to the corresponding author.

\bibliographystyle{mnras}
\bibliography{main}

@ARTICLE{Utsumi2022,
       author = {{Utsumi}, Aoto and {Ohsuga}, Ken and {Takahashi}, Hiroyuki R. and {Asahina}, Yuta},
        title = "{Component of Energy Flow from Supercritical Accretion Disks Around Rotating Stellar Mass Black Holes}",
      journal = {\apj},
         year = 2022,
        month = aug,
       volume = {935},
       number = {1},
          eid = {26},
        pages = {26},
          doi = {10.3847/1538-4357/ac7eb8},
archivePrefix = {arXiv},
       eprint = {2207.02560},
 primaryClass = {astro-ph.HE},
       adsurl = {https://ui.adsabs.harvard.edu/abs/2022ApJ...935...26U}
}

@ARTICLE{Thomsen2022,
       author = {{Thomsen}, Lars L. and {Kwan}, Tom M. and {Dai}, Lixin and {Wu}, Samantha C. and {Roth}, Nathaniel and {Ramirez-Ruiz}, Enrico},
        title = "{Dynamical Unification of Tidal Disruption Events}",
      journal = {\apjl},
         year = 2022,
        month = oct,
       volume = {937},
       number = {2},
          eid = {L28},
        pages = {L28},
          doi = {10.3847/2041-8213/ac911f},
archivePrefix = {arXiv},
       eprint = {2206.02804},
 primaryClass = {astro-ph.HE},
       adsurl = {https://ui.adsabs.harvard.edu/abs/2022ApJ...937L..28T}
}

@ARTICLE{Gezari2021,
       author = {{Gezari}, Suvi},
        title = "{Tidal Disruption Events}",
      journal = {\araa},
         year = 2021,
        month = sep,
       volume = {59},
        pages = {21-58},
          doi = {10.1146/annurev-astro-111720-030029},
archivePrefix = {arXiv},
       eprint = {2104.14580},
 primaryClass = {astro-ph.HE},
       adsurl = {https://ui.adsabs.harvard.edu/abs/2021ARA&A..59...21G}
}

@ARTICLE{Guolo2024,
       author = {{Guolo}, Muryel and {Gezari}, Suvi and {Yao}, Yuhan and {van Velzen}, Sjoert and {Hammerstein}, Erica and {Cenko}, S. Bradley and {Tokayer}, Yarone M.},
        title = "{A Systematic Analysis of the X-Ray Emission in Optically Selected Tidal Disruption Events: Observational Evidence for the Unification of the Optically and X-Ray-selected Populations}",
      journal = {\apj},
         year = 2024,
        month = may,
       volume = {966},
       number = {2},
          eid = {160},
        pages = {160},
          doi = {10.3847/1538-4357/ad2f9f},
archivePrefix = {arXiv},
       eprint = {2308.13019},
 primaryClass = {astro-ph.HE},
       adsurl = {https://ui.adsabs.harvard.edu/abs/2024ApJ...966..160G}
}

@ARTICLE{Curd2021,
       author = {{Curd}, Brandon},
        title = "{Global simulations of tidal disruption event disc formation via stream injection in GRRMHD}",
      journal = {\mnras},
         year = 2021,
        month = nov,
       volume = {507},
       number = {3},
        pages = {3207-3227},
          doi = {10.1093/mnras/stab2172},
archivePrefix = {arXiv},
       eprint = {2105.09904},
 primaryClass = {astro-ph.HE},
       adsurl = {https://ui.adsabs.harvard.edu/abs/2021MNRAS.507.3207C}
}

@ARTICLE{Curd2019,
       author = {{Curd}, Brandon and {Narayan}, Ramesh},
        title = "{GRRMHD simulations of tidal disruption event accretion discs around supermassive black holes: jet formation, spectra, and detectability}",
      journal = {\mnras},
         year = 2019,
        month = feb,
       volume = {483},
       number = {1},
        pages = {565-592},
          doi = {10.1093/mnras/sty3134},
archivePrefix = {arXiv},
       eprint = {1811.06971},
 primaryClass = {astro-ph.HE},
       adsurl = {https://ui.adsabs.harvard.edu/abs/2019MNRAS.483..565C}
}

@ARTICLE{Kato2004,
       author = {{Kato}, Y. and {Mineshige}, S. and {Shibata}, K.},
        title = "{Magnetohydrodynamic Accretion Flows: Formation of Magnetic Tower Jet and Subsequent Quasi-Steady State}",
      journal = {\apj},
         year = 2004,
        month = apr,
       volume = {605},
       number = {1},
        pages = {307-320},
          doi = {10.1086/381234},
archivePrefix = {arXiv},
       eprint = {astro-ph/0307306},
 primaryClass = {astro-ph},
       adsurl = {https://ui.adsabs.harvard.edu/abs/2004ApJ...605..307K}
}

@ARTICLE{Metzger2022,
       author = {{Metzger}, Brian D.},
        title = "{Cooling Envelope Model for Tidal Disruption Events}",
      journal = {\apjl},
         year = 2022,
        month = sep,
       volume = {937},
       number = {1},
          eid = {L12},
        pages = {L12},
          doi = {10.3847/2041-8213/ac90ba},
archivePrefix = {arXiv},
       eprint = {2207.07136},
 primaryClass = {astro-ph.HE},
       adsurl = {https://ui.adsabs.harvard.edu/abs/2022ApJ...937L..12M}
}

@ARTICLE{Ryu2023,
       author = {{Ryu}, Taeho and {Krolik}, Julian and {Piran}, Tsvi and {Noble}, Scott C. and {Avara}, Mark},
        title = "{Shocks Power Tidal Disruption Events}",
      journal = {\apj},
         year = 2023,
        month = nov,
       volume = {957},
       number = {1},
          eid = {12},
        pages = {12},
          doi = {10.3847/1538-4357/acf5de},
archivePrefix = {arXiv},
       eprint = {2305.05333},
 primaryClass = {astro-ph.HE},
       adsurl = {https://ui.adsabs.harvard.edu/abs/2023ApJ...957...12R}
}

@ARTICLE{Sadowski+2015b,
       author = {{S{\k{a}}dowski}, Aleksander and {Narayan}, Ramesh},
        title = "{Powerful radiative jets in supercritical accretion discs around non-spinning black holes}",
      journal = {\mnras},
         year = 2015,
        month = nov,
       volume = {453},
       number = {3},
        pages = {3213-3221},
          doi = {10.1093/mnras/stv1802},
archivePrefix = {arXiv},
       eprint = {1503.00654},
 primaryClass = {astro-ph.HE},
       adsurl = {https://ui.adsabs.harvard.edu/abs/2015MNRAS.453.3213S}
}

@INPROCEEDINGS{2012JPhCS.372a2040T,
       author = {{Tchekhovskoy}, Alexander and {McKinney}, Jonathan C. and {Narayan}, Ramesh},
        title = "{General Relativistic Modeling of Magnetized Jets from Accreting Black Holes}",
    booktitle = {Journal of Physics Conference Series},
         year = 2012,
       series = {Journal of Physics Conference Series},
       volume = {372},
        month = jul,
          eid = {012040},
        pages = {012040},
          doi = {10.1088/1742-6596/372/1/012040},
archivePrefix = {arXiv},
       eprint = {1202.2864},
 primaryClass = {astro-ph.HE},
       adsurl = {https://ui.adsabs.harvard.edu/abs/2012JPhCS.372a2040T}
}

@ARTICLE{2011MNRAS.418L..79T,
       author = {{Tchekhovskoy}, Alexander and {Narayan}, Ramesh and {McKinney}, Jonathan C.},
        title = "{Efficient generation of jets from magnetically arrested accretion on a rapidly spinning black hole}",
      journal = {\mnras},
         year = 2011,
        month = nov,
       volume = {418},
       number = {1},
        pages = {L79-L83},
          doi = {10.1111/j.1745-3933.2011.01147.x},
archivePrefix = {arXiv},
       eprint = {1108.0412},
 primaryClass = {astro-ph.HE},
       adsurl = {https://ui.adsabs.harvard.edu/abs/2011MNRAS.418L..79T}
}

@ARTICLE{Dai2018,
       author = {{Dai}, Lixin and {McKinney}, Jonathan C. and {Roth}, Nathaniel and {Ramirez-Ruiz}, Enrico and {Miller}, M. Coleman},
        title = "{A Unified Model for Tidal Disruption Events}",
      journal = {\apjl},
         year = 2018,
        month = jun,
       volume = {859},
       number = {2},
          eid = {L20},
        pages = {L20},
          doi = {10.3847/2041-8213/aab429},
archivePrefix = {arXiv},
       eprint = {1803.03265},
 primaryClass = {astro-ph.HE},
       adsurl = {https://ui.adsabs.harvard.edu/abs/2018ApJ...859L..20D}
}

@ARTICLE{Steinberg2024,
       author = {{Steinberg}, Elad and {Stone}, Nicholas C.},
        title = "{Stream-disk shocks as the origins of peak light in tidal disruption events}",
      journal = {\nat},
         year = 2024,
        month = jan,
       volume = {625},
       number = {7995},
        pages = {463-467},
          doi = {10.1038/s41586-023-06875-y},
archivePrefix = {arXiv},
       eprint = {2206.10641},
 primaryClass = {astro-ph.HE},
       adsurl = {https://ui.adsabs.harvard.edu/abs/2024Natur.625..463S}
}

@ARTICLE{2013MNRAS.429.3533S,
       author = {{S{\k{a}}dowski}, Aleksander and {Narayan}, Ramesh and {Tchekhovskoy}, Alexander and {Zhu}, Yucong},
        title = "{Semi-implicit scheme for treating radiation under M1 closure in general relativistic conservative fluid dynamics codes}",
      journal = {\mnras},
         year = 2013,
        month = mar,
       volume = {429},
       number = {4},
        pages = {3533-3550},
          doi = {10.1093/mnras/sts632},
archivePrefix = {arXiv},
       eprint = {1212.5050},
 primaryClass = {astro-ph.HE},
       adsurl = {https://ui.adsabs.harvard.edu/abs/2013MNRAS.429.3533S}
}

@ARTICLE{2014MNRAS.439..503S,
       author = {{S{\k{a}}dowski}, Aleksander and {Narayan}, Ramesh and {McKinney}, Jonathan C. and {Tchekhovskoy}, Alexander},
        title = "{Numerical simulations of super-critical black hole accretion flows in general relativity}",
      journal = {\mnras},
         year = 2014,
        month = mar,
       volume = {439},
       number = {1},
        pages = {503-520},
          doi = {10.1093/mnras/stt2479},
archivePrefix = {arXiv},
       eprint = {1311.5900},
 primaryClass = {astro-ph.HE},
       adsurl = {https://ui.adsabs.harvard.edu/abs/2014MNRAS.439..503S}
}

@ARTICLE{2014MNRAS.441.3177M,
       author = {{McKinney}, Jonathan C. and {Tchekhovskoy}, Alexander and {Sadowski}, Aleksander and {Narayan}, Ramesh},
        title = "{Three-dimensional general relativistic radiation magnetohydrodynamical simulation of super-Eddington accretion, using a new code HARMRAD with M1 closure}",
      journal = {\mnras},
         year = 2014,
        month = jul,
       volume = {441},
       number = {4},
        pages = {3177-3208},
          doi = {10.1093/mnras/stu762},
archivePrefix = {arXiv},
       eprint = {1312.6127},
 primaryClass = {astro-ph.CO},
       adsurl = {https://ui.adsabs.harvard.edu/abs/2014MNRAS.441.3177M}
}

@ARTICLE{2015MNRAS.447...49S,
       author = {{S{\k{a}}dowski}, Aleksander and {Narayan}, Ramesh and {Tchekhovskoy}, Alexander and {Abarca}, David and {Zhu}, Yucong and {McKinney}, Jonathan C.},
        title = "{Global simulations of axisymmetric radiative black hole accretion discs in general relativity with a mean-field magnetic dynamo}",
      journal = {\mnras},
         year = 2015,
        month = feb,
       volume = {447},
       number = {1},
        pages = {49-71},
          doi = {10.1093/mnras/stu2387},
archivePrefix = {arXiv},
       eprint = {1407.4421},
 primaryClass = {astro-ph.HE},
       adsurl = {https://ui.adsabs.harvard.edu/abs/2015MNRAS.447...49S}
}

@ARTICLE{2015MNRAS.454.2372S,
       author = {{S{\k{a}}dowski}, Aleksander and {Narayan}, Ramesh},
        title = "{Photon-conserving Comptonization in simulations of accretion discs around black holes}",
      journal = {\mnras},
         year = 2015,
        month = dec,
       volume = {454},
       number = {3},
        pages = {2372-2380},
          doi = {10.1093/mnras/stv2022},
archivePrefix = {arXiv},
       eprint = {1508.04980},
 primaryClass = {astro-ph.HE},
       adsurl = {https://ui.adsabs.harvard.edu/abs/2015MNRAS.454.2372S}
}

@ARTICLE{2016MNRAS.456.3929S,
       author = {{S{\k{a}}dowski}, Aleksander and {Narayan}, Ramesh},
        title = "{Three-dimensional simulations of supercritical black hole accretion discs - luminosities, photon trapping and variability}",
      journal = {\mnras},
         year = 2016,
        month = mar,
       volume = {456},
       number = {4},
        pages = {3929-3947},
          doi = {10.1093/mnras/stv2941},
archivePrefix = {arXiv},
       eprint = {1509.03168},
 primaryClass = {astro-ph.HE},
       adsurl = {https://ui.adsabs.harvard.edu/abs/2016MNRAS.456.3929S}
}

@ARTICLE{Sadowski2016,
       author = {{S{\k{a}}dowski}, Aleksander and {Tejeda}, Emilio and {Gafton}, Emanuel and {Rosswog}, Stephan and {Abarca}, David},
        title = "{Magnetohydrodynamical simulations of a deep tidal disruption in general relativity}",
      journal = {\mnras},
         year = 2016,
        month = jun,
       volume = {458},
       number = {4},
        pages = {4250-4268},
          doi = {10.1093/mnras/stw589},
archivePrefix = {arXiv},
       eprint = {1512.04865},
 primaryClass = {astro-ph.HE},
       adsurl = {https://ui.adsabs.harvard.edu/abs/2016MNRAS.458.4250S}
}

@ARTICLE{2017MNRAS.466..705S,
       author = {{S{\k{a}}dowski}, Aleksander and {Wielgus}, Maciek and {Narayan}, Ramesh and {Abarca}, David and {McKinney}, Jonathan C. and {Chael}, Andrew},
        title = "{Radiative, two-temperature simulations of low-luminosity black hole accretion flows in general relativity}",
      journal = {\mnras},
         year = 2017,
        month = apr,
       volume = {466},
       number = {1},
        pages = {705-725},
          doi = {10.1093/mnras/stw3116},
archivePrefix = {arXiv},
       eprint = {1605.03184},
 primaryClass = {astro-ph.HE},
       adsurl = {https://ui.adsabs.harvard.edu/abs/2017MNRAS.466..705S}
}

@ARTICLE{2003ApJ...589..444G,
       author = {{Gammie}, Charles F. and {McKinney}, Jonathan C. and {T{\'o}th}, G{\'a}bor},
        title = "{HARM: A Numerical Scheme for General Relativistic Magnetohydrodynamics}",
      journal = {\apj},
         year = 2003,
        month = may,
       volume = {589},
       number = {1},
        pages = {444-457},
          doi = {10.1086/374594},
archivePrefix = {arXiv},
       eprint = {astro-ph/0301509},
 primaryClass = {astro-ph},
       adsurl = {https://ui.adsabs.harvard.edu/abs/2003ApJ...589..444G}
}

@ARTICLE{2022ApJ...935L...1L,
       author = {{Liska}, M.~T.~P. and {Musoke}, G. and {Tchekhovskoy}, A. and {Porth}, O. and {Beloborodov}, A.~M.},
        title = "{Formation of Magnetically Truncated Accretion Disks in 3D Radiation-transport Two-temperature GRMHD Simulations}",
      journal = {\apjl},
         year = 2022,
        month = aug,
       volume = {935},
       number = {1},
          eid = {L1},
        pages = {L1},
          doi = {10.3847/2041-8213/ac84db},
archivePrefix = {arXiv},
       eprint = {2201.03526},
 primaryClass = {astro-ph.HE},
       adsurl = {https://ui.adsabs.harvard.edu/abs/2022ApJ...935L...1L}
}

@ARTICLE{1993ApJS...88..253S,
       author = {{Sutherland}, Ralph S. and {Dopita}, M.~A.},
        title = "{Cooling Functions for Low-Density Astrophysical Plasmas}",
      journal = {\apjs},
         year = 1993,
        month = sep,
       volume = {88},
        pages = {253},
          doi = {10.1086/191823},
       adsurl = {https://ui.adsabs.harvard.edu/abs/1993ApJS...88..253S}
}

@ARTICLE{2017MNRAS.467.3604R,
       author = {{Ressler}, S.~M. and {Tchekhovskoy}, A. and {Quataert}, E. and {Gammie}, C.~F.},
        title = "{The disc-jet symbiosis emerges: modelling the emission of Sagittarius A* with electron thermodynamics}",
      journal = {\mnras},
         year = 2017,
        month = may,
       volume = {467},
       number = {3},
        pages = {3604-3619},
          doi = {10.1093/mnras/stx364},
archivePrefix = {arXiv},
       eprint = {1611.09365},
 primaryClass = {astro-ph.HE},
       adsurl = {https://ui.adsabs.harvard.edu/abs/2017MNRAS.467.3604R}
}

@INPROCEEDINGS{1973blho.conf..343N,
       author = {{Novikov}, I.~D. and {Thorne}, K.~S.},
        title = "{Astrophysics of black holes.}",
    booktitle = {Black Holes (Les Astres Occlus)},
         year = 1973,
        month = jan,
        pages = {343-450},
       adsurl = {https://ui.adsabs.harvard.edu/abs/1973blho.conf..343N}
}

@ARTICLE{1995ApJ...438L..37A,
       author = {{Abramowicz}, Marek A. and {Chen}, Xingming and {Kato}, Shoji and {Lasota}, Jean-Pierre and {Regev}, Oded},
        title = "{Thermal Equilibria of Accretion Disks}",
      journal = {\apjl},
         year = 1995,
        month = jan,
       volume = {438},
        pages = {L37},
          doi = {10.1086/187709},
archivePrefix = {arXiv},
       eprint = {astro-ph/9409018},
 primaryClass = {astro-ph},
       adsurl = {https://ui.adsabs.harvard.edu/abs/1995ApJ...438L..37A}
}

@ARTICLE{2011ApJ...732...52Z,
       author = {{Zheng}, Sheng-Ming and {Yuan}, Feng and {Gu}, Wei-Min and {Lu}, Ju-Fu},
        title = "{Revisiting the Thermal Stability of Radiation-dominated Thin Disks}",
      journal = {\apj},
         year = 2011,
        month = may,
       volume = {732},
       number = {1},
          eid = {52},
        pages = {52},
          doi = {10.1088/0004-637X/732/1/52},
archivePrefix = {arXiv},
       eprint = {1103.0347},
 primaryClass = {astro-ph.HE},
       adsurl = {https://ui.adsabs.harvard.edu/abs/2011ApJ...732...52Z}
}

@ARTICLE{2014ApJ...786....6L,
       author = {{Li}, Shuang-Liang and {Begelman}, Mitchell C.},
        title = "{Thermal Stability of a Thin Disk with Magnetically Driven Winds}",
      journal = {\apj},
         year = 2014,
        month = may,
       volume = {786},
       number = {1},
          eid = {6},
        pages = {6},
          doi = {10.1088/0004-637X/786/1/6},
archivePrefix = {arXiv},
       eprint = {1403.2305},
 primaryClass = {astro-ph.HE},
       adsurl = {https://ui.adsabs.harvard.edu/abs/2014ApJ...786....6L}
}

@ARTICLE{2019ApJ...887..256H,
       author = {{Habibi}, Asiyeh and {Abbassi}, Shahram},
        title = "{Thermal Instability of Thin Accretion Disks in the Presence of Wind and a Toroidal Magnetic Field}",
      journal = {\apj},
         year = 2019,
        month = dec,
       volume = {887},
       number = {2},
          eid = {256},
        pages = {256},
          doi = {10.3847/1538-4357/ab5793},
archivePrefix = {arXiv},
       eprint = {1911.06645},
 primaryClass = {astro-ph.GA},
       adsurl = {https://ui.adsabs.harvard.edu/abs/2019ApJ...887..256H}
}

@ARTICLE{2022ApJ...930..108W,
       author = {{Wu}, Wen-Biao and {Gu}, Wei-Min and {Sun}, Mouyuan},
        title = "{Thermal Equilibrium Solutions of Black Hole Accretion Flows: Outflows versus Advection}",
      journal = {\apj},
         year = 2022,
        month = may,
       volume = {930},
       number = {2},
          eid = {108},
        pages = {108},
          doi = {10.3847/1538-4357/ac6588},
archivePrefix = {arXiv},
       eprint = {2204.03606},
 primaryClass = {astro-ph.HE},
       adsurl = {https://ui.adsabs.harvard.edu/abs/2022ApJ...930..108W}
}

@ARTICLE{2023ApJ...954..150H,
       author = {{Huang}, Jiahui and {Feng}, Hua and {Gu}, Wei-Min and {Wu}, Wen-Biao},
        title = "{Black Hole Accretion with Saturated Magnetic Pressure and Disk Wind}",
      journal = {\apj},
         year = 2023,
        month = sep,
       volume = {954},
       number = {2},
          eid = {150},
        pages = {150},
          doi = {10.3847/1538-4357/ace71e},
archivePrefix = {arXiv},
       eprint = {2307.06585},
 primaryClass = {astro-ph.HE},
       adsurl = {https://ui.adsabs.harvard.edu/abs/2023ApJ...954..150H}
}

@ARTICLE{2021ApJ...919L..20D,
       author = {{Dexter}, Jason and {Scepi}, Nicolas and {Begelman}, Mitchell C.},
        title = "{Radiation GRMHD Simulations of the Hard State of Black Hole X-Ray Binaries and the Collapse of a Hot Accretion Flow}",
      journal = {\apjl},
         year = 2021,
        month = oct,
       volume = {919},
       number = {2},
          eid = {L20},
        pages = {L20},
          doi = {10.3847/2041-8213/ac2608},
archivePrefix = {arXiv},
       eprint = {2109.06239},
 primaryClass = {astro-ph.HE},
       adsurl = {https://ui.adsabs.harvard.edu/abs/2021ApJ...919L..20D}
}

@ARTICLE{Curd2023,
       author = {{Curd}, Brandon and {Narayan}, Ramesh},
        title = "{GRRMHD simulations of MAD accretion discs declining from super-Eddington to sub-Eddington accretion rates}",
      journal = {\mnras},
         year = 2023,
        month = jan,
       volume = {518},
       number = {3},
        pages = {3441-3461},
          doi = {10.1093/mnras/stac3330},
archivePrefix = {arXiv},
       eprint = {2209.12081},
 primaryClass = {astro-ph.HE},
       adsurl = {https://ui.adsabs.harvard.edu/abs/2023MNRAS.518.3441C}
}

@ARTICLE{2024ApJ...966...47L,
       author = {{Liska}, M.~T.~P. and {Kaaz}, N. and {Chatterjee}, K. and {Emami}, Razieh and {Musoke}, G.},
        title = "{Magnetic Flux Plays an Important Role during a Black Hole X-Ray Binary Outburst in Radiative Two-temperature General Relativistic Magnetohydrodynamic Simulations}",
      journal = {\apj},
         year = 2024,
        month = may,
       volume = {966},
       number = {1},
          eid = {47},
        pages = {47},
          doi = {10.3847/1538-4357/ad344a},
archivePrefix = {arXiv},
       eprint = {2309.15926},
 primaryClass = {astro-ph.HE},
       adsurl = {https://ui.adsabs.harvard.edu/abs/2024ApJ...966...47L}
}

@ARTICLE{Meza2025,
       author = {{Meza}, Maria Renee and {Huang}, Xiaoshan and {Davis}, Shane W. and {Jiang}, Yan-Fei},
        title = "{Radiation-magnetohydrodynamic Simulations of Accretion Flow Formation After a Tidal Disruption Event}",
      journal = {\apj},
         year = 2025,
        month = nov,
       volume = {993},
       number = {1},
          eid = {57},
        pages = {57},
          doi = {10.3847/1538-4357/ae044c},
archivePrefix = {arXiv},
       eprint = {2506.00109},
 primaryClass = {astro-ph.HE},
       adsurl = {https://ui.adsabs.harvard.edu/abs/2025ApJ...993...57M}
}

@ARTICLE{Hills1975,
       author = {{Hills}, J.~G.},
        title = "{Possible power source of Seyfert galaxies and QSOs}",
      journal = {\nat},
         year = 1975,
        month = mar,
       volume = {254},
       number = {5498},
        pages = {295-298},
          doi = {10.1038/254295a0},
       adsurl = {https://ui.adsabs.harvard.edu/abs/1975Natur.254..295H}
}

@ARTICLE{Rees1988,
       author = {{Rees}, Martin J.},
        title = "{Tidal disruption of stars by black holes of {}10$^{6}$-{}10$^{8}$ solar masses in nearby galaxies}",
      journal = {\nat},
         year = 1988,
        month = jun,
       volume = {333},
       number = {6173},
        pages = {523-528},
          doi = {10.1038/333523a0},
       adsurl = {https://ui.adsabs.harvard.edu/abs/1988Natur.333..523R}
}

@INPROCEEDINGS{Phinney1989,
       author = {{Phinney}, E.~S.},
        title = "{Manifestations of a Massive Black Hole in the Galactic Center}",
    booktitle = {The Center of the Galaxy},
         year = 1989,
       editor = {{Morris}, Mark},
       volume = {136},
        month = jan,
        pages = {543},
        series = {},
       adsurl = {https://ui.adsabs.harvard.edu/abs/1989IAUS..136..543P}
}

@ARTICLE{Evans1989,
       author = {{Evans}, Charles R. and {Kochanek}, Christopher S.},
        title = "{The Tidal Disruption of a Star by a Massive Black Hole}",
      journal = {\apjl},
         year = 1989,
        month = nov,
       volume = {346},
        pages = {L13},
          doi = {10.1086/185567},
       adsurl = {https://ui.adsabs.harvard.edu/abs/1989ApJ...346L..13E}
}

@ARTICLE{2013ApJ...767...25G,
       author = {{Guillochon}, James and {Ramirez-Ruiz}, Enrico},
        title = "{Hydrodynamical Simulations to Determine the Feeding Rate of Black Holes by the Tidal Disruption of Stars: The Importance of the Impact Parameter and Stellar Structure}",
      journal = {\apj},
         year = 2013,
        month = apr,
       volume = {767},
       number = {1},
          eid = {25},
        pages = {25},
          doi = {10.1088/0004-637X/767/1/25},
archivePrefix = {arXiv},
       eprint = {1206.2350},
 primaryClass = {astro-ph.HE},
       adsurl = {https://ui.adsabs.harvard.edu/abs/2013ApJ...767...25G}
}

@ARTICLE{2017A&A...600A.124M,
       author = {{Mainetti}, Deborah and {Lupi}, Alessandro and {Campana}, Sergio and {Colpi}, Monica and {Coughlin}, Eric R. and {Guillochon}, James and {Ramirez-Ruiz}, Enrico},
        title = "{The fine line between total and partial tidal disruption events}",
      journal = {\aap},
         year = 2017,
        month = apr,
       volume = {600},
          eid = {A124},
        pages = {A124},
          doi = {10.1051/0004-6361/201630092},
archivePrefix = {arXiv},
       eprint = {1702.07730},
 primaryClass = {astro-ph.HE},
       adsurl = {https://ui.adsabs.harvard.edu/abs/2017A&A...600A.124M}
}

@ARTICLE{2019ApJ...882L..26G,
       author = {{Golightly}, E.~C.~A. and {Nixon}, C.~J. and {Coughlin}, E.~R.},
        title = "{On the Diversity of Fallback Rates from Tidal Disruption Events with Accurate Stellar Structure}",
      journal = {\apjl},
         year = 2019,
        month = sep,
       volume = {882},
       number = {2},
          eid = {L26},
        pages = {L26},
          doi = {10.3847/2041-8213/ab380d},
archivePrefix = {arXiv},
       eprint = {1907.05895},
 primaryClass = {astro-ph.HE},
       adsurl = {https://ui.adsabs.harvard.edu/abs/2019ApJ...882L..26G}
}

@ARTICLE{2025ApJ...983..177N,
       author = {{Navarro Navarro}, N{\'u}ria and {Piran}, Tsvi},
        title = "{Once a Giant, (Almost) Always a Giant: Partial Tidal Disruption Events of Giant Stars}",
      journal = {\apj},
         year = 2025,
        month = apr,
       volume = {983},
       number = {2},
          eid = {177},
        pages = {177},
          doi = {10.3847/1538-4357/ad96b7},
archivePrefix = {arXiv},
       eprint = {2411.15346},
 primaryClass = {astro-ph.HE},
       adsurl = {https://ui.adsabs.harvard.edu/abs/2025ApJ...983..177N}
}

@ARTICLE{Zauderer2013,
       author = {{Zauderer}, B.~A. and {Berger}, E. and {Margutti}, R. and {Pooley}, G.~G. and {Sari}, R. and {Soderberg}, A.~M. and {Brunthaler}, A. and {Bietenholz}, M.~F.},
        title = "{Radio Monitoring of the Tidal Disruption Event Swift J164449.3+573451. II. The Relativistic Jet Shuts Off and a Transition to Forward Shock X-Ray/Radio Emission}",
      journal = {\apj},
         year = 2013,
        month = apr,
       volume = {767},
       number = {2},
          eid = {152},
        pages = {152},
          doi = {10.1088/0004-637X/767/2/152},
archivePrefix = {arXiv},
       eprint = {1212.1173},
 primaryClass = {astro-ph.HE},
       adsurl = {https://ui.adsabs.harvard.edu/abs/2013ApJ...767..152Z}
}

@ARTICLE{Pasham2015,
       author = {{Pasham}, Dheeraj R. and {Cenko}, S. Bradley and {Levan}, Andrew J. and {Bower}, Geoffrey C. and {Horesh}, Assaf and {Brown}, Gregory C. and {Dolan}, Stephen and {Wiersema}, Klaas and {Filippenko}, Alexei V. and {Fruchter}, Andrew S. and {Greiner}, Jochen and {O'Brien}, Paul T. and {Page}, Kim L. and {Rau}, Arne and {Tanvir}, Nial R.},
        title = "{A Multiwavelength Study of the Relativistic Tidal Disruption Candidate Swift J2058.4+0516 at Late Times}",
      journal = {\apj},
         year = 2015,
        month = may,
       volume = {805},
       number = {1},
          eid = {68},
        pages = {68},
          doi = {10.1088/0004-637X/805/1/68},
archivePrefix = {arXiv},
       eprint = {1502.01345},
 primaryClass = {astro-ph.HE},
       adsurl = {https://ui.adsabs.harvard.edu/abs/2015ApJ...805...68P}
}

@ARTICLE{2022ApJ...937....8Y,
       author = {{Yao}, Yuhan and {Lu}, Wenbin and {Guolo}, Muryel and {Pasham}, Dheeraj R. and {Gezari}, Suvi and {Gilfanov}, Marat and {Gendreau}, Keith C. and {Harrison}, Fiona and {Cenko}, S. Bradley and {Kulkarni}, S.~R. and {Miller}, Jon M. and {Walton}, Dominic J. and {Garc{\'\i}a}, Javier A. and {van Velzen}, Sjoert and {Alexander}, Kate D. and {Miller-Jones}, James C.~A. and {Nicholl}, Matt and {Hammerstein}, Erica and {Medvedev}, Pavel and {Stern}, Daniel and {Ravi}, Vikram and {Sunyaev}, R. and {Bloom}, Joshua S. and {Graham}, Matthew J. and {Kool}, Erik C. and {Mahabal}, Ashish A. and {Masci}, Frank J. and {Purdum}, Josiah and {Rusholme}, Ben and {Sharma}, Yashvi and {Smith}, Roger and {Sollerman}, Jesper},
        title = "{The Tidal Disruption Event AT2021ehb: Evidence of Relativistic Disk Reflection, and Rapid Evolution of the Disk-Corona System}",
      journal = {\apj},
         year = 2022,
        month = sep,
       volume = {937},
       number = {1},
          eid = {8},
        pages = {8},
          doi = {10.3847/1538-4357/ac898a},
archivePrefix = {arXiv},
       eprint = {2206.12713},
 primaryClass = {astro-ph.HE},
       adsurl = {https://ui.adsabs.harvard.edu/abs/2022ApJ...937....8Y}
}

@ARTICLE{Eftekhari2024,
       author = {{Eftekhari}, T. and {Tchekhovskoy}, A. and {Alexander}, K.~D. and {Berger}, E. and {Chornock}, R. and {Laskar}, T. and {Margutti}, R. and {Yao}, Y. and {Cendes}, Y. and {Gomez}, S. and {Hajela}, A. and {Pasham}, D.~R.},
        title = "{Late-time X-Ray Observations of the Jetted Tidal Disruption Event AT2022cmc: The Relativistic Jet Shuts Off}",
      journal = {\apj},
         year = 2024,
        month = oct,
       volume = {974},
       number = {2},
          eid = {149},
        pages = {149},
          doi = {10.3847/1538-4357/ad72ea},
archivePrefix = {arXiv},
       eprint = {2404.10036},
 primaryClass = {astro-ph.HE},
       adsurl = {https://ui.adsabs.harvard.edu/abs/2024ApJ...974..149E}
}

@ARTICLE{Lancova2019,
       author = {{Lan{\v{c}}ov{\'a}}, Debora and {Abarca}, David and {Klu{\'z}niak}, W{\l}odek and {Wielgus}, Maciek and {S{\k{a}}dowski}, Aleksander and {Narayan}, Ramesh and {Schee}, Jan and {T{\"o}r{\"o}k}, Gabriel and {Abramowicz}, Marek},
        title = "{Puffy Accretion Disks: Sub-Eddington, Optically Thick, and Stable}",
      journal = {\apjl},
         year = 2019,
        month = oct,
       volume = {884},
       number = {2},
          eid = {L37},
        pages = {L37},
          doi = {10.3847/2041-8213/ab48f5},
archivePrefix = {arXiv},
       eprint = {1908.08396},
 primaryClass = {astro-ph.HE},
       adsurl = {https://ui.adsabs.harvard.edu/abs/2019ApJ...884L..37L}
}

@ARTICLE{Narayan2003,
       author = {{Narayan}, Ramesh and {Igumenshchev}, Igor V. and {Abramowicz}, Marek A.},
        title = "{Magnetically Arrested Disk: an Energetically Efficient Accretion Flow}",
      journal = {\pasj},
         year = 2003,
        month = dec,
       volume = {55},
        pages = {L69-L72},
          doi = {10.1093/pasj/55.6.L69},
archivePrefix = {arXiv},
       eprint = {astro-ph/0305029},
 primaryClass = {astro-ph},
       adsurl = {https://ui.adsabs.harvard.edu/abs/2003PASJ...55L..69N}
}

@ARTICLE{2025MNRAS.539.3473Q,
       author = {{Qiao}, Erlin and {Wu}, Yongxin and {Lin}, Yiyang and {Guo}, Meng and {Liu}, Jifeng and {Guo}, Chenlei and {Jin}, Chichuan and {Jiang}, Ning},
        title = "{Early evolution of super-Eddington accretion flow in tidal disruption events}",
      journal = {\mnras},
         year = 2025,
        month = jun,
       volume = {539},
       number = {4},
        pages = {3473-3488},
          doi = {10.1093/mnras/staf719},
archivePrefix = {arXiv},
       eprint = {2505.02434},
 primaryClass = {astro-ph.HE},
       adsurl = {https://ui.adsabs.harvard.edu/abs/2025MNRAS.539.3473Q}
}

@ARTICLE{2017MNRAS.470.2367C,
       author = {{Chael}, Andrew A. and {Narayan}, Ramesh and {Sadowski}, Aleksander},
        title = "{Evolving non-thermal electrons in simulations of black hole accretion}",
      journal = {\mnras},
         year = 2017,
        month = sep,
       volume = {470},
       number = {2},
        pages = {2367-2386},
          doi = {10.1093/mnras/stx1345},
archivePrefix = {arXiv},
       eprint = {1704.05092},
 primaryClass = {astro-ph.HE},
       adsurl = {https://ui.adsabs.harvard.edu/abs/2017MNRAS.470.2367C}
}

@ARTICLE{2025ApJ...993...14A,
       author = {{Alush}, Yael and {Stone}, Nicholas C.},
        title = "{Late-time Evolution of Magnetized Disks in Tidal Disruption Events}",
      journal = {\apj},
         year = 2025,
        month = nov,
       volume = {993},
       number = {1},
          eid = {14},
        pages = {14},
          doi = {10.3847/1538-4357/adf217},
archivePrefix = {arXiv},
       eprint = {2503.03811},
 primaryClass = {astro-ph.HE},
       adsurl = {https://ui.adsabs.harvard.edu/abs/2025ApJ...993...14A}
}

@ARTICLE{2025ApJ...985...77P,
       author = {{Piro}, Anthony L. and {Mockler}, Brenna},
        title = "{Late-time Evolution and Instabilities of Tidal Disruption Disks}",
      journal = {\apj},
         year = 2025,
        month = may,
       volume = {985},
       number = {1},
          eid = {77},
        pages = {77},
          doi = {10.3847/1538-4357/adc729},
archivePrefix = {arXiv},
       eprint = {2412.01922},
 primaryClass = {astro-ph.HE},
       adsurl = {https://ui.adsabs.harvard.edu/abs/2025ApJ...985...77P}
}

@ARTICLE{2025ApJ...989...27T,
       author = {{Tuna}, Semih and {Metzger}, Brian D. and {Jiang}, Yan-Fei and {White}, Christopher},
        title = "{Time-dependent Radiation Transport Simulations of Infrared Echoes from Dust-shrouded Luminous Transients}",
      journal = {\apj},
         year = 2025,
        month = aug,
       volume = {989},
       number = {1},
          eid = {27},
        pages = {27},
          doi = {10.3847/1538-4357/ade8ed},
archivePrefix = {arXiv},
       eprint = {2501.13157},
 primaryClass = {astro-ph.HE},
       adsurl = {https://ui.adsabs.harvard.edu/abs/2025ApJ...989...27T}
}

@ARTICLE{1999ApJ...522..839W,
       author = {{Wang}, Jian-Min and {Szuszkiewicz}, Ewa and {Lu}, Fang-Jun and {Zhou}, You-Yuan},
        title = "{Emergent Spectra from Slim Accretion Disks in Active Galactic Nuclei}",
      journal = {\apj},
         year = 1999,
        month = sep,
       volume = {522},
       number = {2},
        pages = {839-845},
          doi = {10.1086/307686},
       adsurl = {https://ui.adsabs.harvard.edu/abs/1999ApJ...522..839W}
}

@ARTICLE{2025ApJ...981..144R,
       author = {{Roth}, Nathaniel and {Anninos}, Peter and {Fragile}, P. Chris and {Pickrel}, Derrick},
        title = "{X-Ray Spectra from General Relativistic Radiation Magnetohydrodynamic Simulations of Thin Disks}",
      journal = {\apj},
         year = 2025,
        month = mar,
       volume = {981},
       number = {2},
          eid = {144},
        pages = {144},
          doi = {10.3847/1538-4357/adb1c1},
archivePrefix = {arXiv},
       eprint = {2501.18040},
 primaryClass = {astro-ph.HE},
       adsurl = {https://ui.adsabs.harvard.edu/abs/2025ApJ...981..144R}
}

@ARTICLE{2021NatAs...5..491H,
       author = {{Horesh}, A. and {Cenko}, S.~B. and {Arcavi}, I.},
        title = "{Delayed radio flares from a tidal disruption event}",
      journal = {Nature Astronomy},
         year = 2021,
        month = may,
       volume = {5},
        pages = {491-497},
          doi = {10.1038/s41550-021-01300-8},
archivePrefix = {arXiv},
       eprint = {2102.11290},
 primaryClass = {astro-ph.HE},
       adsurl = {https://ui.adsabs.harvard.edu/abs/2021NatAs...5..491H}
}

@ARTICLE{2024ApJ...971..185C,
       author = {{Cendes}, Y. and {Berger}, E. and {Alexander}, K.~D. and {Chornock}, R. and {Margutti}, R. and {Metzger}, B. and {Wieringa}, M.~H. and {Bietenholz}, M.~F. and {Hajela}, A. and {Laskar}, T. and {Stroh}, M.~C. and {Terreran}, G.},
        title = "{Ubiquitous Late Radio Emission from Tidal Disruption Events}",
      journal = {\apj},
         year = 2024,
        month = aug,
       volume = {971},
       number = {2},
          eid = {185},
        pages = {185},
          doi = {10.3847/1538-4357/ad5541},
archivePrefix = {arXiv},
       eprint = {2308.13595},
 primaryClass = {astro-ph.HE},
       adsurl = {https://ui.adsabs.harvard.edu/abs/2024ApJ...971..185C}
}

@ARTICLE{Curd2023b,
       author = {{Curd}, Brandon and {Emami}, Razieh and {Anantua}, Richard and {Palumbo}, Daniel and {Doeleman}, Sheperd and {Narayan}, Ramesh},
        title = "{Jets from SANE super-Eddington accretion discs: morphology, spectra, and their potential as targets for ngEHT}",
      journal = {\mnras},
         year = 2023,
        month = feb,
       volume = {519},
       number = {2},
        pages = {2812-2837},
          doi = {10.1093/mnras/stac3716},
archivePrefix = {arXiv},
       eprint = {2206.06358},
 primaryClass = {astro-ph.HE},
       adsurl = {https://ui.adsabs.harvard.edu/abs/2023MNRAS.519.2812C}
}

@ARTICLE{2020MNRAS.492..686L,
       author = {{Lu}, Wenbin and {Bonnerot}, Cl{\'e}ment},
        title = "{Self-intersection of the fallback stream in tidal disruption events}",
      journal = {\mnras},
         year = 2020,
        month = feb,
       volume = {492},
       number = {1},
        pages = {686-707},
          doi = {10.1093/mnras/stz3405},
archivePrefix = {arXiv},
       eprint = {1904.12018},
 primaryClass = {astro-ph.HE},
       adsurl = {https://ui.adsabs.harvard.edu/abs/2020MNRAS.492..686L}
}

@ARTICLE{2025ApJ...988L..24H,
       author = {{Hu}, Fangyi (Fitz) and {Goodwin}, Adelle and {Price}, Daniel J. and {Mandel}, Ilya and {Sari}, Re'em and {Hayasaki}, Kimitake},
        title = "{Radio Emission from Tidal Disruption Events Produced by the Collision between Super-Eddington Outflows and the Circumnuclear Medium}",
      journal = {\apjl},
         year = 2025,
        month = jul,
       volume = {988},
       number = {1},
          eid = {L24},
        pages = {L24},
          doi = {10.3847/2041-8213/adeb79},
archivePrefix = {arXiv},
       eprint = {2507.01273},
 primaryClass = {astro-ph.HE},
       adsurl = {https://ui.adsabs.harvard.edu/abs/2025ApJ...988L..24H}
}

@ARTICLE{2026ApJ..1001...71A,
       author = {{Abolmasov}, Pavel and {Bromberg}, Omer and {Levinson}, Amir and {Nakar}, Ehud},
        title = "{Tidal Disruption of a Magnetized Star}",
      journal = {\apj},
         year = 2026,
        month = apr,
       volume = {1001},
       number = {1},
          eid = {71},
        pages = {71},
          doi = {10.3847/1538-4357/ae4ebc},
archivePrefix = {arXiv},
       eprint = {2509.23894},
 primaryClass = {astro-ph.HE},
       adsurl = {https://ui.adsabs.harvard.edu/abs/2026ApJ..1001...71A}
}

@ARTICLE{2016MNRAS.461.3760H,
       author = {{Hayasaki}, Kimitake and {Stone}, Nicholas and {Loeb}, Abraham},
        title = "{Circularization of tidally disrupted stars around spinning supermassive black holes}",
      journal = {\mnras},
         year = 2016,
        month = oct,
       volume = {461},
       number = {4},
        pages = {3760-3780},
          doi = {10.1093/mnras/stw1387},
archivePrefix = {arXiv},
       eprint = {1501.05207},
 primaryClass = {astro-ph.HE},
       adsurl = {https://ui.adsabs.harvard.edu/abs/2016MNRAS.461.3760H}
}

@ARTICLE{2016MNRAS.455.2253B,
       author = {{Bonnerot}, Cl{\'e}ment and {Rossi}, Elena M. and {Lodato}, Giuseppe and {Price}, Daniel J.},
        title = "{Disc formation from tidal disruptions of stars on eccentric orbits by Schwarzschild black holes}",
      journal = {\mnras},
         year = 2016,
        month = jan,
       volume = {455},
       number = {2},
        pages = {2253-2266},
          doi = {10.1093/mnras/stv2411},
archivePrefix = {arXiv},
       eprint = {1501.04635},
 primaryClass = {astro-ph.HE},
       adsurl = {https://ui.adsabs.harvard.edu/abs/2016MNRAS.455.2253B}
}

\label{lastpage}
\end{document}